 \documentstyle[aps,twocolumn]{revtex}            
\def\zt{\tilde{z}}
\def\half{\mbox{$\frac{1}{2}$}}     

\begin{document}
\preprint{UMD PP-98-69}
\draft
\title{Stable 3-level leapfrog integration in numerical relativity}
\author{Kimberly C. B. New}
\address{Department of Physics \& Atmospheric Science, 
Drexel University, Philadelphia 19104}
\author{Keith Watt}
\address{Department of Astronomy, University of Maryland, College
Park 20742-2421}
\author{Charles W. Misner}
\address{Department of Physics, University of Maryland, College Park
20742-4111}
\author{Joan M. Centrella}
\address{Department of Physics \& Atmospheric Science, 
Drexel University, Philadelphia 19104}

\date{{\it Physical Review D}, in press}
\maketitle
    \begin{abstract}
    The 3-level leapfrog time integration algorithm is an attractive choice
for numerical relativity simulations since it is time-symmetric and
avoids non-physical damping.
    In Newtonian problems without velocity
dependent forces, this method enjoys the advantage of long term
stability.
    However, for more general differential equations, whether
ordinary or partial,
delayed onset numerical instabilities can arise and destroy the solution.
    A known cure for such instabilities
appears to have been overlooked in
many application areas.  
    We give an improved cure (``deloused
leapfrog'') that both reduces memory demands (important for $3+1$
dimensional wave equations) and allows for the use of adaptive timesteps
without a loss in accuracy.
    We
show both that the instability arises and that the cure we propose works in
highly relativistic problems such as tightly bound geodesics, spatially
homogeneous spacetimes, and strong gravitational waves.
    In the gravitational wave test case
(polarized waves in a Gowdy spacetime) the deloused
leapfrog method was five to eight times less CPU costly at various accuracies
than the implicit Crank-Nicholson method, which is not subject to this
instability.
    \end{abstract} 

\section{INTRODUCTION}

Numerical relativity comprises the dynamical solution of the Einstein
equations on a computer, allowing the construction of spacetimes that
cannot be studied by purely analytic methods.  
    A major application of numerical relativity is the modeling of
astrophysical sources of gravitational radiation such as binary black
hole \cite{alliance_prl} or neutron star inspiral \cite{mw}, and
nonspherical stellar collapse \cite{ps}.  
    The continued development of gravitational wave detectors, with
the expectation that ground-based interferometers such as LIGO
\cite{ligo}, VIRGO \cite{virgo} and GEO600 \cite{geo} will begin
taking data in a few years, gives these studies a high priority.  
    Numerical relativity is also important for studying the dynamics
of pure gravitational waves \cite{anninos-waves}, inhomogeneous
cosmologies \cite{k-sc}, the behavior of
cosmological singularities \cite{singularity,bm93}, and critical behavior
in general relativity \cite{critical}.

All of these endeavors require accurate numerical algorithms to
correctly model the physics of curved spacetime.
    Simulations in three spatial dimensions plus time are expensive in
terms of both CPU usage and memory requirements, and thus demand
numerical methods that are efficient in both these regards.
    Memory limits, however, are less elastic in the short term than
CPU time constraints.
    Thus a three-level second order algorithm may be more appropriate
than a faster, high order algorithm which can only be implemented on
smaller problems.
    Also, modeling the inspiral of binary black holes or neutron
stars requires evolving the system for many orbital periods, so that
numerical algorithms with long term stability and freedom from
unphysical damping are essential.

Leapfrog methods are often used for the time integration of equations
in numerical relativity and other branches of computational physics.
    The 3-level leapfrog method has the important property of being
symplectic.  
    In the context of a Hamiltonian system for which
the differential equation has a symplectic structure
(conjugate pairing of coordinates and momenta), this means that the difference
equations also have such a structure and the integration step in the
difference equations is a canonical transformation.
    With a symplectic integrator, all the Lagrangian integral
invariants, including phase space volume, are exactly conserved by
the integration scheme.
    Since the leapfrog method is time symmetric and maintains good
conservation of physically conserved quantities \cite{ruth,hut,QKSL},
it has a well-deserved reputation in the context of Newtonian
mechanics. 
    Unfortunately this reputation is generally not merited when velocity
dependent forces are met.
    In the integration of systems with such forces,
this scheme is well-known to be susceptible to numerical
instability (e.g., \cite{np59,bf73,ss85,aa89,caa93}, and references
therein), even under conditions where local linearization analysis
anticipates stability.
    This instability occurs in the integration of both ordinary and
partial differential equations and, in the case of partial
differential equations, is independent of the mesh size used for the
spatial discretization \cite{ss85}.

    An understanding of the origin of this instability was given by
Sanz-Serna \cite{ss85} who pointed out that the leapfrog scheme
approximates not merely the intended differential equation system but
a larger ``augmented'' system containing additional, nonphysical,
parasitic modes.
    Since the leapfrog method is symplectic as applied to the augmented
system \cite{ssv87}, the advantages of symplectic methods (see
\cite{ruth}) would be attained to the extent that the parasitic modes
remain zero numerically, as they do in an exact solution of the
augmented system \cite{wm}.
    Aoyagi and Abe \cite{aa89,aa91} identified the diagnostic symptom
of this instability as a sawtooth oscillation or alternation of values
between odd and even steps of the integration, and supplied a
cure---a Runge-Kutta smoother to suppress this oscillation.
    Subsequent work \cite{caa93} shows these phenomena in
ordinary differential equations, where the delayed onset of this
instability is clearly apparent.

We have studied the use of the 3-level leapfrog method in numerical
relativity.
    In this work (see also \cite{wm}), we extend the ideas of Aoyagi
and collaborators for removing the unstable parasitic modes,
yielding an algorithm that reduces the number of time levels of data
that must be stored by the code, allows the timestep to be changed,
and thus is better suited to long-term integration of large scale
numerical systems. 
    Although we concentrate on the ADM 3+1 formalism for numerical
relativity, our methods are quite general and thereby applicable to a
wide range of problems in computational physics.
    Section~\ref{instability} describes the source of the 3-level
leapfrog instability. 
    Section~\ref{deloused} presents our algorithm, dubbed ``deloused
leapfrog,'' for removing instabilities that may arise in 3-level
leapfrog integrations.  
    In Sec.~\ref{3+1}, we summarize the $3+1$ formalism of numerical
relativity and in Sec.~\ref{gowdy} introduce three simple classes of
Gowdy $T^3$ spacetimes. 
    In Sec.~\ref{sims}, we use these models as relativistic testbeds
for the deloused leapfrog technique and evaluate the numerical
efficiency of deloused leapfrog by comparing its cost-effectiveness
with those of the staggered leapfrog and Crank-Nicholson techniques. 
    A summary and discussion of the results of the stability and
efficiency tests of the deloused leapfrog technique
is given in Sec.~\ref{conclusions}.

\section{The 3-level Leapfrog Instability} \label{instability}

We begin by considering the system of differential equations
    \begin{equation}
    \frac{d{\bf z}}{dt} = {\bf F}({\bf z},t).
    \label{ode}
    \end{equation} 
    Although we use a system of {\em ordinary} differential
equations [Eqs.\ (\ref{ode})] to describe the leapfrog instability
and its cure in this section and the next, the techniques we outline
apply equally well to systems of {\em partial} differential equations
in which the time integration is carried out using the leapfrog
method; see Sec.~\ref{sims} below.
    The 3-level leapfrog discretization of this system is
    \begin{equation}
    {\bf \zt}^{n+2} = {\bf \zt}^{n} + 2 {\bf F}({\bf
\zt}^{n+1},t^{n+1}) \Delta t ,
    \label{discrete-ode}
    \end{equation} 
    where we assume a constant timestep $\Delta t$ between time
levels $n$ and $n + 1$. 
    The distinction between ${\bf z}$ in the differential equation
and ${\bf \zt}$ in the difference equation is a warning that the
relationship is not as straighforward as first appears.
    The discretization given in Eqs.\ (\ref{discrete-ode}) has ${\cal
O}({\Delta t}^2)$ accuracy and is a 3-level method, in that knowledge
of data on time levels $n$ and $n+1$ is needed to compute the result
on time level $n+2$. 
    The 3-level leapfrog algorithm [Eqs.\ (\ref{discrete-ode})] is
symplectic and time reversible, which means that it provides a 
Hamiltonian (damping free) model of an underlying Hamiltonian
differential equation system \cite{ruth,QKSL}.

Equations (\ref{ode}) arise in many physical applications in which ${\bf
z}$ comprises both position and velocity data ${\bf z} = (\vec r,
\vec v)$.
    For example, the motion of a particle of mass $\mu$ under the
action of a force $\vec f = \mu \vec a$ is given by the set of
equations
    \begin{eqnarray}
    \frac{d{\vec r}}{dt} & = & \vec v \nonumber \\
    \frac{d{\vec v}}{dt} & = & \vec a ,
    \label{newt}
    \end{eqnarray} 
    where $\vec r$ is the position vector of the particle.  
    The 3-level leapfrog discretization of Eqs.\ (\ref{newt}) gives
    \begin{eqnarray}
    \vec r\,^{n+1} & = & \vec r\,^{n-1} + 2 \vec v\,^n \Delta t 
\nonumber \\
    \vec v\,^{n+1} & = & \vec v\,^{n-1} + 2 \vec a\,^n \Delta t \quad .
    \label{discrete-newt}
    \end{eqnarray} 
    For a particle moving in a Newtonian gravitational field
$\vec{a}\,^n = \vec{a}\,^n(\vec{r}\,^n)$ and the system given by
Eqs.\ (\ref{discrete-newt}) is stable and thus suitable for long time
integrations.  
    However, if there are so-called ``velocity dependent forces'' in
which ${\vec a}\,^n = {\vec a}\,^{n}({\vec r}\,^n,{\vec v}\,^n)$, such
as arise for a particle moving under the influence of a magnetic
field or a general relativistic gravitational field, the 3-level
leapfrog scheme [Eqs.\ (\ref{discrete-newt})] can become unstable.  
    As we shall explain below, nonphysical parasitic modes can arise
during the time integration and eventually destroy the numerical
solution.

Notice that the leapfrog algorithm [Eqs.\ (\ref{discrete-ode})] gives
the value of ${\bf \zt}^{n+2}$  at, say, the even time level $n+2$ in
terms of the value ${\bf \zt}^{n}$ at the even level $n$ and the
source term ${\bf F}({\bf \zt}^{n+1},t^{n+1})$ at the odd level $n+1$.  
    It also requires the specification of initial data at two time
levels, ${\bf \zt}^0$ and ${\bf \zt}^1$, which is twice as much as the
original first-order differential equation system requires.  
    This doubling of initial condition specifications is the clue to
the fact that this numerical algorithm using ${\bf \zt}$ has twice as
many degrees of freedom as does the physical system where states are
specified by ${\bf z}$.
    The algorithm can be expressed, with a change of notation, by
writing the solution at the even timesteps as ${\bf \zt}^{2n} = 
{\bf x}^{2n}$ and at the odd timesteps as ${\bf \zt}^{2n+1} = 
{\bf y}^{2n+1}$.  
    With this, Eqs.\ (\ref{discrete-ode}) become
    \begin{eqnarray}
    {\bf x}^{2n+2} & = & {\bf x}^{2n} +
    2 {\bf F}({\bf y}^{2n+1},t^{2n+1}) \Delta t  \nonumber \\
    {\bf y}^{2n+3} & = & {\bf y}^{2n + 1} +
    2 {\bf F}({\bf x}^{2n+2},t^{2n+2}) \Delta t \quad .
    \label{augmented}
    \end{eqnarray} 
    Sanz-Serna \cite{ss85} (concisely summarized in \cite{vss86})
notes that Eqs.\ (\ref{augmented}) can be considered to be a
consistent single-step discretization of a larger system of equations
for the even solutions ${\bf x}$ and the odd solutions ${\bf y}$,
    \begin{eqnarray}
    \frac{d{\bf x}}{dt} & = & {\bf F}({\bf y},t) \nonumber \\
    \frac{d{\bf y}}{dt} & = & {\bf F}({\bf x},t),
    \label{augmented-ode}
    \end{eqnarray} 
    with timestep $2 \Delta t$.  
    Equations (\ref{augmented-ode}) are known as the augmented system
\cite{ss85,vss86}.

    If ${\bf z}$ is a solution of Eqs.~(\ref{ode}), it gives a
solution of Eqs.~(\ref{augmented-ode}), the augmented system, as 
${\bf x}={\bf z}$ and ${\bf y}={\bf z}$.  
    In general, however, other (unphysical) solutions will be
possible.  
    Since these unphysical, parasitic solutions can arise as valid solutions to
the augmented difference system of Eqs.~(\ref{augmented}), the
3-level leapfrog integrator alone cannot distinguish between them and
the physical solutions.  

Since the physical solutions of Eqs.~(\ref{augmented-ode}) have $\bf
x = y$ it is natural to define 
    \begin{equation}\label{parasitic-def}
        \bf x = z + w \quad , \quad y = z - w \quad ,
    \end{equation}
    so that the $\bf w$ measure the parasitic deviations from the
desired physical solutions $\bf z$.
    One can rewrite Eqs.~(\ref{parasitic-def}) in finite difference form,
using the definitions preceding Eqs.~(\ref{augmented}), as
    \begin{equation}\label{z-x-y}
    {\bf \zt}^n = {\bf z}^n + (-1)^n {\bf w}^n \quad .
    \end{equation}
    The solution ${\bf \zt}^n$ to the leapfrog difference system [Eqs.\ (\ref
{discrete-ode})] thus contains both physical ${\bf z}^n$ and
parasitic modes ${\bf w}^n$.
    With this notation, the augmented differential equations can be written
as
\cite{caa93}
    \begin{eqnarray}
    2\frac{d{\bf z}}{dt} & = & {\bf F}({\bf z-w},t) + {\bf F}({\bf z+w},t)
                        \label{constraint-ode}  \quad ,\\
    2\frac{d{\bf w}}{dt} & = & {\bf F}({\bf z-w},t) - {\bf F}({\bf z+w},t)
    \nonumber           \quad .
    \end{eqnarray} 
    We note that when these are expanded in powers of ${\bf w}$ one has
    \begin{eqnarray}
    \frac{d{\bf z}}{dt} &  = & {\bf F}({\bf z},t) + {\cal O} ({\bf w}^2)
                        \label{parametric-ode}  \quad ,\\
    \frac{d{\bf w}}{dt} &  = & - {\bf DF}({\bf z},t)\cdot {\bf w}
                                + {\cal O} ({\bf w}^3)
    \nonumber
    \end{eqnarray} 
    where ${\bf DF}$ is the matrix of partial derivatives
$\partial F_i/\partial z_j$.

    The second of Eqs.~(\ref{parametric-ode}), ignoring the cubic and
higher terms, is a linear equation (in ${\bf w}$) which from the time
dependence in ${{\bf DF}({\bf z}(t),t)}$ easily gives rise to parametric
amplification \cite{LL} leading to growth of the parasitic modes
${\bf w}$.
    Such parametric amplification was clearly diagnosed in a useful
example \cite{aa89}.
    A still stronger argument for the growth of parasitic modes
(without the above linear perturbation assumptions) had been given
earlier by Sanz-Serna \cite{ss85,vss86} based on his proof that the
leapfrog scheme preserves volume in the augmented state space even
when the original system is not Hamiltonian.
    Sanz-Serna has an elementary example, $dz/dt = z^2$, $z(0) = -1$,
with a two dimensional $zw$ augmented phase space (Fig.~1 in
\cite{ss85}); this shows all the qualitative features of the delayed
onset instability typical of numerical catastrophies that generally
result from using leapfrog in nonlinear systems with velocity dependent forces.
    In this example, {\em every} solution diverges ($z \rightarrow
\infty$) if at any point $w \neq  0$, although  $z \rightarrow 0$ as
$t \rightarrow \infty$ for the physical $w = 0$ solution.
    Numerical experiments with leapfrog follow the divergent solutions
of this analytical example since $w \neq  0$ at some point arises
either from imperfect initial conditions, roundoff error, or
discretization error.
    Hence, nonlinear interaction or parametric amplification of linear
interaction between the physical and parasitic modes of the solution to the 
leapfrog difference system [Eqs.\ (\ref {discrete-ode})] can cause the
parasitic mode to grow to the point where it destroys the numerical solution.

    However, if the augmented difference system
[Eqs.~(\ref{discrete-ode}) or (\ref{augmented})]
decouples, no such instability occurs.
    To illustrate what it means for these equations to decouple,
consider the augmented difference system for Eqs.~(\ref{discrete-newt}),
in the absence of velocity dependent forces.
    In that case one can 
let ${\bf x} = (\vec r,\vec u)$ and ${\bf y} = (\vec q,\vec v)$ to
find that the (even $\vec r$, odd $\vec v$) system does not couple to
the (odd $\vec q$, even $\vec u$) system.
    Thus the augmented difference system
consists of two interlaced, noninteracting
copies (even $\vec r$, odd $\vec v$) and (odd $\vec q$, even $\vec
u$) of the physical system.
    Each of these two systems is of the Newtonian form where the
leapfrog scheme has shown itself remarkably stable.

    Thus, as mentioned above, in a 3-level leapfrog integration of
Eqs.~(\ref{newt}) stablity can generally only be anticipated in the absence of
velocity dependent forces.
    In this case, the parasitic mode
is generally still present as the difference between two interlaced
numerical solutions of Eqs.~(\ref{newt}), but will remain small unless
the physical system is highly sensitive to small differences in
initial conditions (i.e., chaotic).
    Note that it is customary, in the absence of velocity dependent
forces, to omit the $\vec q$ and $\vec u$
variables from the integration scheme, yielding the staggered
leapfrog scheme (see below).
    Alternatively, in a code based on Eqs.~(\ref{discrete-ode}), one
could in this case solve for two independent solutions approximating
Eqs.~(\ref{ode}) based on distinct initial conditions for $(\vec r,
\vec v)$ and for $(\vec q, \vec u)$.

    A stable 3-level leapfrog integration of a system of equations containing
``velocity dependent forces'' can be maintained if the growth of the
parasitic mode can be controlled.
    When a constraint $\bf w = 0$ is adjoined to
Eqs.~(\ref{constraint-ode}) they reduce to the original physical
system of Eq.~(\ref{ode}).
    This constraint, when imposed initially, is preserved by
the differential equation system (\ref{constraint-ode}) since $\bf
w=0$ gives $d{\bf w}/dt = 0$.
    But in numerical implementations errors will inevitably introduce
nonzero $\bf w$.  
    The cure proposed by Aoyagi and Abe \cite{aa89,aa91}
is to reimpose the constraint $\bf w=0$ as necessary to suppress the
parasitic mode.
    Their method for doing so is based on the identification of 
${\bf w}$ via its signature even-odd timepstep alternation in sign,
which is evident in Eq.~(\ref{z-x-y}).
    As noted in \cite{wm}, this method is very efficient since, from
the power series expansions in Eqs.~(\ref{parametric-ode}), the
parasitic modes ${\bf w}$ need merely be suppressed to single
precision accuracy to assure that they do not contaminate the
physical solution ${\bf z}$ in double precision.
     We will return to the cure in Sec.~\ref{deloused} below.

Note that the instability under discussion here is not related to the
``mesh-drifting'' instability inherent to non-dissipative leapfrog
integrations of partial differential equations \cite{nr}.  
    As has been emphasized above, the 3-level leapfrog instability
can arise in integrations of both ordinary and partial differential
equations and results from the temporal, not spatial, discretization
of the method.

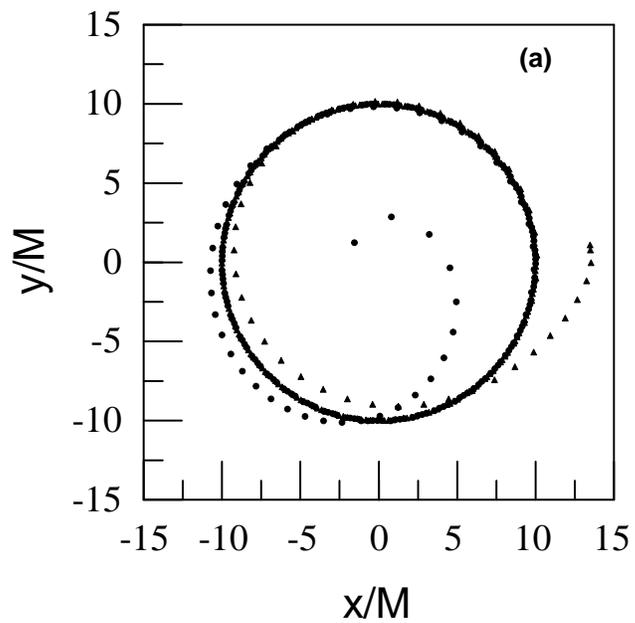
\begin{figure}
    \caption[Numerical integration, geodesic, 3-level leapfrog.]{The
numerical integration of the geodesic equations for a particle in the
Schwarzschild spacetime using the 3-level leapfrog technique.  
    The geodesic equation was solved in rectangular coordinates from
a 3D Hamiltonian (see \cite{wm}).
    The particle was given initial conditions such that it should
remain on a circular orbit of radius $r_0 = 10 M$ and have an orbit
period of $199 M$.  
    In each frame, the data is plotted on every twenty-third timestep
($\Delta t=0.1 M$). 
    The instability manifests as the solutions on odd (circles) and
even (triangles) timesteps diverge.
    Although the integrator appears to be stable at early times, the
parasitic mode is present from the beginning, on a much smaller scale
than is used in these plots. 
    The integrator failed and the code crashed after $\sim 6$ orbital
periods.  
    $(a)$ The particle orbit in the $x$-$y$ plane. 
    $(b)$ The magnitude of the particle's position vector as a
function of time.
    \label{unstable-orbit}}
    \end{figure}

To demonstrate this instability, we have used the 3-level leapfrog
method to numerically integrate the geodesic equations for a particle
moving on a circular orbit of radius $r_0$ in the Schwarzschild
spacetime.  
    In the limit $r_0 \gg M$, we recover the usual Newtonian
equations of motion [Eqs.~(\ref{newt})].  
    In this case, the leapfrog method produced a circular orbit that
was stable for ten thousand orbital periods (before we terminated the
run).
    However, as $r_0$ is decreased, general relativistic effects give
rise to terms that behave like velocity-dependent forces and the
integrator fails to maintain a stable evolution.  
    Fig.~\ref{unstable-orbit} shows the results of the geodesic
integration for the case $r_0 = 10 M$.
    The particle orbit shown in Fig.~\ref{unstable-orbit}a initially
appears to be stable; eventually the instability manifests as the
solutions on the even and odd timesteps diverge.  
    Fig.~\ref{unstable-orbit}b shows the magnitude of the particle's
position vector as a function of time; again the even and odd
solutions clearly diverge as the parasitic mode grows to destructive
levels.
    Although the parasitic mode is present from the beginning of the
calculation, it takes a number of orbits before it grows to
noticeable levels; the instability then grows catastrophically,
causing the integration to crash after about six orbits.

Two other integration methods (whose efficiency we will compare with
that of our deloused leapfrog algorithm in Sec.~\ref{sims}),
staggered leapfrog and Crank-Nicholson, do not suffer from the
instability present in the 3-level leapfrog algorithm.
    In both cases this happens because neither of these algorithms
augments the phase space (or state space) of the problem by adding
new degrees of freedom not found in the physical system.
    To illustrate this, we will continue to use the integration of 
Eqs.~(\ref{newt}) as the example upon which we base our outline of
these integration algorithms.  

    As mentioned above, the even and
odd degrees of freedom in Eqs.~(\ref{newt}) can
be written ${\bf x} = (\vec r,\vec u)$ and ${\bf y} = (\vec q,\vec v)$. 
    But when the force law lets $\vec{a} = \vec{a}(\vec{r})$ be
calculated independently of $\vec v$, the $(\vec r,\vec v)$ pair of
variables are not coupled to the $(\vec q,\vec u)$ pair, so this
second pair can be dropped from the numerical algorithm.
    This leads to the methodology of the staggered leapfrog algorithm, which
defines the
variables it evolves on alternating time levels only. 
    A staggered leapfrog integration of Eqs.~(\ref{newt}) would, for example,
evaluate only $\vec r$ at even steps and only $\vec v$ at odd steps.
    It is then customary to renumber the steps so that
the even steps are integer values, the odd half integer:
    \begin{eqnarray}
    \vec r\,^{n+1}   & = & \vec r\,^{n} + \vec v\,^{n+1/2} \Delta t  
    \nonumber \\
    \vec v\,^{n+3/2} & = & \vec v\,^{n+1/2} + \vec a\,^{n+1} \Delta t 
                              \quad ,
    \label{staggered}
    \end{eqnarray} 
    where the constant $\Delta t$ is the difference between two
consecutive integer (or half-integer) time levels \cite{nr}.  
    The initial conditions are specified by giving $(\vec r\,^0,
\vec v\,^{1/2})$.
    This method is time-symmetric and symplectic and thus avoids 
nonphysical damping \cite{ruth,hut}.
    Here, with velocity independent Newtonian forces, one has a 2-level
method.
    [I.e., only the $\vec v$ components are used in updating the
$\vec r$'s, and only the $\vec r$ components in updating the 
$\vec v$'s.]
    This leapfrog method gives second order accuracy
at the same computational cost as the first order Euler method.

    When $\vec a$ depends on $\vec v$ as well as $\vec r$, a method
such as extrapolation (e.g., $\vec{v}\,^{n+1} =
(3/2)\vec{v}\,^{n+1/2} - (1/2)\vec{v}\,^{n-1/2}$) is needed to
calculate $\vec v$ at the integer (or even) levels; this generally
destroys the time-symmetric nature of the algorithm and leads to
errors in conserved quantities.

The Crank-Nicholson technique is an iterative integration algorithm
\cite{gko}.  
    For the system given in Eqs.\ (\ref{newt}), the iteration cycle
is initialized by setting
    \begin{equation}
    \vec r\,^{n+1}=\vec r\,^{n},\ \vec v\,^{n+1}=\vec v\,^{n}.
    \label{cn-init}
    \end{equation} 
    Then data at the half integer levels $n+1$ is determined via
    \begin{eqnarray}
    \vec r\,^{n+1/2}&=&\frac{1}{2}(\vec r\,^{n+1}+\vec r\,^{n})
\nonumber \\
    \vec v\,^{n+1/2}&=&\frac{1}{2}(\vec v\,^{n+1}+\vec v\,^{n}) \quad.
    \label{cn1}
    \end{eqnarray} 
    Next, the values of the data at level $n+1$ are updated according
to
    \begin{eqnarray}
    \vec r\,^{n+1}&=&\vec r\,^{n}+\vec v\,^{n+1/2} \Delta t  \nonumber
\\
    \vec v\,^{n+1}&=&\vec v\,^{n}+\vec a\,^{n+1/2} \Delta t \quad,
    \label{cn2}
    \end{eqnarray} 
    where $\vec a\,^{n+1/2}$ has been computed based on the data
given by Eqs.\ (\ref{cn1}).  
    The steps given by Eqs.\ (\ref{cn1}) and (\ref{cn2}) are then
repeated until the relative differences of the values of each of the
components of $\vec r\,^{n+1}$ and $\vec v\,^{n+1}$ between adjacent
iteration cycles have each converged to the desired level of
accuracy.
    This method requires the specification of initial data $\vec
r\,^0$ and $\vec v\,^0$.  
    It has accuracy ${\cal O}({\Delta t}^2)$; it is time symmetric if
the iteration converges to machine precision.

Both the staggered leapfrog and the Crank-Nicholson algorithms avoid
the instability arising from an augmented phase space.
    One sees in each case that there is a single physical pair of
values at each timestep.
    These are $(\vec r\,^{n}, \vec v\,^{n+1/2})$ for staggered
leapfrog and $(\vec r\,^{n}, \vec v\,^{n})$ for Crank-Nicholson.
    Other quantities such as $\vec v\,^{n+1}$ in staggered leapfrog
or the half integer values in Crank-Nicholson are temporary variables
computed from the physical data.
    Thus no extraneous degrees of freedom, like the $\bf w$ quantities
in 3-level leapfrog, appear.

\section{Deloused leapfrog} \label{deloused}

The instability in the 3-level leapfrog scheme is ideally eliminated
with a technique that retains as much of the symplectic character of
the 3-level scheme as possible.  
    Aoyagi and Abe \cite{aa89,aa91} give such a prescription to
remove the parasitic modes before they grow large enough to destroy
the calculation.  
    This method relies on the Runge-Kutta (RK) method \cite{nr} and is
called the RK smoother.  
    (The usual RK algorithm is fourth order; however,
Aoyagi and Abe do not state what order RK scheme is used in their smoother.)
    They compare it to a less effective second
order smoother suggested by a colleague.
    We have extended the work of Aoyagi and Abe to produce a second
order algorithm which requires less storage and also allows the use
of adaptive (i.e., not constant in time) timesteps $\Delta t$ (see
also \cite{wm}). 
    We term the 3-level method with this improved smoother ``deloused
leapfrog,'' since it removes the parasitic modes from the
calculation.
    In this section, we give the algorithm for this deloused leapfrog
method.  
   Ref.~\cite{wm} discusses its properties in Hamiltonian problems
with applications to highly noncircular and highly relativistic
geodesics.
   Those applications use the adaptive timestep our delousing routine
allows.

The prescription for a delousing step to remove the parasitic modes
uses a second order Runga-Kutta (RK2) algorithm. 
    (The usual RK algorithm is fourth order.) 
    Our algorithm proceeds as follows.  
    Referring to the 3-level discretization in
Eq.~(\ref{discrete-newt}), we assume that data is available on (say)
an odd level $n$ and an even level $n-1$. 
    First, use RK2 to evolve the data a half-step backward from the
odd level $n$ and a half-step forward from the even level $n-1$:
     \begin{description}
        \item[Step (1)] use RK2 with $\delta t =
     -\frac{1}{2}\Delta t$ \\
     to get $\vec r\,^{n-1/2}_{odd}, \vec v\,^{n-1/2}_{odd}$ 
     from $\vec r\,^n, \vec v\,^n$
        \item[Step (2)] use RK2 with $\delta t =
     +\frac{1}{2}\Delta t$ \\
     to get $\vec r\,^{n-1/2}_{even}, \vec v\,^{n-1/2}_{even}$ 
     from $\vec r\,^{n-1}, \vec v\,^{n-1}$,
     \end{description}
     where, for example,
 $\vec r\,^{n-1/2}_{odd}$ represents data on level \mbox{$n-\frac{1}{2}$}
that comes from the odd level $n$.
    We now have data from both odd and even solutions at the same time
level \mbox{$n-\frac{1}{2}$}. 
    Eqs.~(\ref{parasitic-def}) are then utilized to obtain the data $\vec
r\,^{n-1/2}$ and $\vec v\,^{n-1/2}$ that contain only the physical mode:
     \begin{description}
    \item[Step (3)]  set $\vec r\,^{n-1/2} = \frac{1}{2}
    (\vec r\,^{n-1/2}_{even} + \vec r\,^{n-1/2}_{odd})$, \\
    $\vec v\,^{n-1/2} = \frac{1}{2}
    (\vec v\,^{n-1/2}_{even} + \vec v\,^{n-1/2}_{odd})$.
     \end{description}
    With the parasitic mode thus eliminated at level $n-\frac{1}{2}$
(i.e., with the constraint ${\bf w} = 0$ enforced),
we now use RK2 to step the data a half-step forward and backward:
     \begin{description}
    \item[Step (4)] use RK2 with $\delta t^{\prime} =
    +\frac{1}{2}\Delta t^{\prime}$ \\
    to get $\vec r\,^{n}, \vec v\,^{n}$ from
    $\vec r\,^{n-1/2}, \vec v\,^{n-1/2}$
    \item[Step (5)] use RK2 with $\delta t^{\prime} =
    -\frac{1}{2}\Delta t^{\prime}$ \\ 
    to get $\vec r\,^{n-1}, \vec v\,^{n-1}$ from
    $\vec r\,^{n-1/2}, \vec v\,^{n-1/2}$ .
     \end{description}
    Since RK2 does not suffer from the leapfrog instability (for the
same reason as that given in the preceeding section for the
stability of the staggered leapfrog and Crank-Nicholson methods), no
parasitic mode is introduced by its use.  
    With this ``deloused'' data at levels $n$ and $n-1$, which
contain only the physical modes, the 3-level leapfrog integration is
resumed.
    This prescription can be coded so that no more than three levels
of data for $\vec r$ and $\vec v$ need be kept in memory at any one time,
which is an important consideration when it is applied to $3+1$
dimensional partial differential equations.

Steps (4) and (5) of this procedure can also be used to start the
integration. 
    With initial data specified (e.g., at $t=0$) as
values for $\vec r\,^{1/2}, \vec v\,^{1/2}$, this procedure
provides $\vec r\,^{0}, \vec v\,^{0}$ and $\vec r\,^{1}, \vec
v\,^{1}$ to start the 3-level algorithm of Eqs.~(\ref{discrete-ode}).

  Notice that the timesteps $\pm \frac{1}{2} \Delta t^{\prime}$ used
in Steps~(4) and (5) to bring the deloused data from level $n -
\frac{1}{2}$ to levels $n$ and $n-1$ are not required to be the same
as the timesteps $\pm \frac{1}{2} \Delta t$ used in Steps~(1) and
(2).
    This is an improvement on the method of Aoyagi and Abe
\cite{aa89,aa91} and allows the system timestep to be changed
whenever delousing is performed,
with the integration restarted in a time symmetric
manner.  
    In this paper, all our runs were carried out using constant
timesteps; see, however, Ref.~\cite{wm} for an example of adaptive
timesteps applied to highly eccentric particle orbits.

    The deloused leapfrog method is not strictly symplectic because
the delousing steps are taken using RK2, which is not a symplectic
method, and the suppression Step~(3) does not preserve volume in
the augmented phase space.
    Since delousing steps can in principle be rare, 
and ideally do not disturb the
physical solution (when the ${\cal O} ({\bf w}^2)$ terms in
Eqs.~(\ref{parametric-ode}) are beyond machine precision), the
failure here of the time symmetric and symplectic properties
possessed by Eqs.~(\ref{discrete-ode}) could have negligible impact.

To determine when a delousing step is needed, the growth of the
parasitic mode is monitored.  
    Since Eq.~(\ref{parasitic-def}) shows that this parasitic mode
changes sign on each time level $n$, we look for this alternate
timestep oscillation in some quantity that characterizes the system.
    The detection of such an oscillation in the quantity monitored
triggers a call to the delousing module, which then removes the
parasitic mode.
    The goal is to monitor a quantity that manifests the instability
at detectable levels early enough, so that it can be eliminated
before it dominates the integration.
    The best quantity to use for the delousing trigger varies from
problem to problem and also depends on what level of parasitic mode
one is willing to tolerate in the solution (see further discussion in
Secs.\ \ref{sims} and \ref{conclusions}).

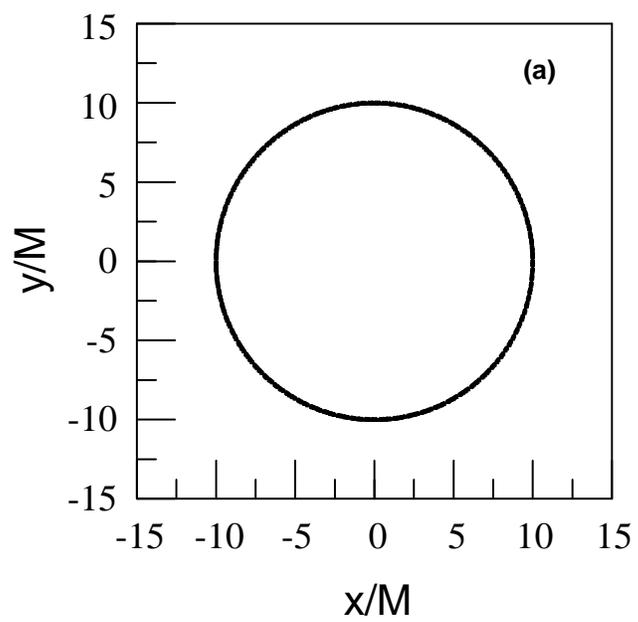
\begin{figure}
    \caption[Numerical integration, geodesic, deloused leapfrog.]{
Same as Fig.~\protect{\ref{unstable-orbit}} except that the
integration was carried out using the deloused leapfrog method.  
    The orbit is now stable for the $\sim 10,000$-orbit duration of
the simulation.  
    The timestep was the same as that used in
Fig.~\protect{\ref{unstable-orbit}}, but the data is plotted only
every 2001 timesteps.  
    Note that even though only $\sim 1$ point per orbit (each orbit
is aproximately 2000 timesteps) are shown, there are over 10,000 
points plotted in the figure.  
    Both odd and even solutions lie directly on top of each other,
filling out the orbit track.  
    The delousing module was applied on average once every 405
timesteps, hence this figure was produced at a 2 percent increase in
CPU cost per orbit over the simulation shown in
Fig.~\protect{\ref{unstable-orbit}}.
    \label{stable-orbit}}
    \end{figure}

Figure \ref{stable-orbit} shows the results of our deloused leapfrog
integration of the same particle equations of motion as in
Fig.~\ref{unstable-orbit}. 
    The delousing trigger chosen in this case was an oscillation in a
quantity based upon the action $dJ = \frac{1}{2}(p\,dq - q\,dp)$ of
Hamiltonian mechanics:
    \begin{eqnarray}
    \Delta J & = & \half(\vec p\,^{n}+\vec p\,^{n-1})\cdot
    (\vec r\,^{n} - \vec r\,^{n-1})
    \nonumber\\
    & & \mbox{\quad}
    - \half(\vec r\,^{n} + \vec r\,^{n-1})\cdot
    (\vec p\,^{n} - \vec p\,^{n-1})
    \nonumber\\
    & = & 
    {\vec r\,^{n}\cdot \vec p\,^{n-1}}-
    {\vec p\,^{n}\cdot \vec r\,^{n-1}}, 
    \label{action}
    \end{eqnarray}
    where $\vec r$ and $\vec p$ are the particle's position and
momentum vectors, respectively. 
    The use of this trigger initiated a delousing step about once
every 405 timesteps.
    The orbit was stable for the duration of our simulation, about
ten thousand orbits.

\section{The $3+1$ formalism of numerical relativity} \label{3+1}

The development of the deloused leapfrog algorithm described in the
preceeding section was motivated by our search for a stable and
efficient time integration technique that could be utilized in CPU
and memory intensive, general relativistic simulations of the orbital
stability of binary neutron stars.  This section briefly summarizes the $3+1$
form of the Einstein equations upon which such numerical relativity
simulations are based.
    Here we consider only vacuum spacetimes; however, our results
apply to models with sources as well.

The Einstein field equations of general relativity are $G_{\mu\nu}=
8\pi T_{\mu\nu}$, where $G_{\mu\nu}$ and $T_{\mu\nu}$ are the
Einstein and stress energy tensors, respectively \cite{MTW}.
    We set $T_{\mu\nu}=0$ (as is the case
for vacuum spacetimes), $G=c=1$ and our convention is such that Greek
indices run from 0 to 3, Latin indices run from 1 to 3, and repeated
raised and lowered indices are summed over.  
    The numerical solution of the Einstein equations is facilitated
by the ADM decomposition \cite{MTW,adm}, which divides four
dimensional spacetime into a series of three dimensional, spacelike
slices that are connected by one dimensional, timelike curves.  
    This ``3+1'' split transforms the Einstein equations into two
sets:  the evolution equations, which govern temporal changes in the
gravitational field variables, and the constraint equations, which
the field variables must satisfy on each spacelike slice.

The general form of the metric in the ADM formalism is
    \begin{equation}
{ds}^2=-(\alpha{dt})^2+g_{ij}({dx}^i+\beta^i{dt})({dx}^j+\beta^j{dt}).
    \label{metric-adm}
    \end{equation} 
    The vacuum evolution equations for the three-metric $g_{ij}$ and
the extrinsic curvature $K_{ij}$ are (see, e.g., \cite{seidel} and
references therein)
    \begin{equation}
    g_{ij,t} = -2 \alpha K_{ij} + D_i \beta_j + D_j \beta_i,
\label{gdot-vacuum}
    \end{equation} 
    and
    \begin{eqnarray}
    K_{ij,t} & = &-D_{i}D_{j}\alpha + \beta^{l}D_{l}K_{ij} +
K_{il}D_{j}\beta^{l}
    +K_{lj}D_{i}\beta^{l} \nonumber \\   &  & 
    +\alpha[R_{ij}-2K_{il}K^{l}_{j}+KK_{ij}],
    \label{kdot-vacuum}
    \end{eqnarray} 
    where $t$ is coordinate time, commas denote partial derivatives,
$K=K^{i}_{i}$, and $D_i$ and
$R_{ij}$ are the covariant derivative and Ricci tensor, respectively,
formed from the three-metric $g_{ij}$.  
    The lapse $\alpha$ and the shift vector $\beta^i$ are freely
specifiable and define the coordinate conditions under which $g_{ij}$
and $K_{ij}$ are evolved.
    The right hand side of Eq.\ (\ref{gdot-vacuum}) may be a function
of $g_{ij}$, as dependence on $g_{ij}$ arises through the covariant
derivative operation when $\beta^i$ is nonzero or when $\alpha$ is
chosen to depend on $g_{ij}$; moreover, the right hand side of Eq.\
(\ref{kdot-vacuum}) is always a function of $K_{ij}$.  
    Thus ``velocity dependent forces'' are generally present in this
system of equations and therefore 3-level leapfrog integrations of
them will suffer from the instablity described in Sec.~II.

The set of constraint equations is comprised of the Hamiltonian
constraint:
    \begin{equation}
    R-K_{ij}K^{ij}+K^2=0, \label{hc-vacuum}
    \end{equation} 
    where $R=R^{i}_{i}$, and momentum constraints:
    \begin{equation}
    D_{j}(K^{ij}-g^{ij}K)=0. \label{mc-vacuum}
    \end{equation}

When the field variables and coordinate conditions for the spacetime
are functions of time only, the Einstein equations [Eqs.\
(\ref{gdot-vacuum}) and (\ref{kdot-vacuum})] governing the evolution
of vacuum spacetimes become ordinary differential equations:
    \begin{equation}
    \frac{dg_{ij}}{dt} = -2 \alpha K_{ij}, 
    \label{gdot-ode}
    \end{equation}
    \begin{equation}
    \frac{dK_{ij}}{dt} =\alpha(KK_{ij}-2K_{il}K^{l}_{j}).
    \label{kdot-ode}
    \end{equation} 
    This simplified set of equations is the basis of the toy
codes discussed in Sec.\ \ref{toy}.

\section{Gowdy $T^3$ spacetimes} \label{gowdy}

In this section, we introduce the Gowdy $T^3$ spacetimes and then
describe three classes of these spacetimes, for which exact evolution
solutions are known and which we have used as relativistic testbeds
for the deloused leapfrog technique (see Sec.\ VI.{}).

The Gowdy $T^3$ spacetimes are solutions of the vacuum Einstein
equations [Eqs.\ (\ref{gdot-vacuum}) and (\ref{kdot-vacuum})] on the
3-torus, in which plane gravitational waves are contained within an
expanding universe \cite{gowdy}.
    Solutions of Gowdy's equations have been used for studying the
nature of the initial cosmological singularity \cite{singularity,bm93},
and, as is the case in this work, for testing numerical codes for
solving Einstein's equations \cite{vanputten}.

The Gowdy $T^3$ metric can be written:
    \begin{equation}
    {ds}^2=e^{(\lambda-\tau)/2}(-e^{2\tau}{d\tau}^{2}+{dz}^2)+
    e^{\tau}{dw}^2, \label{metric-origgowdy}
    \end{equation} 
    where
    \begin{equation}
    {dw}^2=e^P({dx}+Q\,{dy})^2+e^{-P}{dy}^2. 
    \label{dw-origgowdy}
    \end{equation} 
    This form that appears in \cite{bm93} is there attributed to Moncrief
\cite{vm81}, whose exploitation of the harmonic map \cite{es64,cm78}
character of the associated Einstein equations could suggest many
equivalent forms.
    The Gowdy coordinates $\tau$, $\lambda$, $\sigma$, $\delta$, and
$\theta$ from \cite{bm93} have here been written as $-\tau$, $-\lambda$,
$x$, $y$, and $z$.  
    The change in the signs of the time coordinate $\tau$ and of
$\lambda$ signify that time increases as the universe expands; in
most earlier work, the emphasis on the study of the initial
singularity dictated a time coordinate that increased as the universe
contracted.
    The metric parameters $\lambda$, $P$, and $Q$ are functions of
$z$ and $\tau$ only and are periodic in $z$.

We found the use of an alternate time variable
    \begin{equation}
    t=e^{\tau} \label{new-timecdt}
    \end{equation} 
    convenient in that it here makes the (coordinate) velocity of
propagation of gravitational waves constant.
    This means that with a fixed spatial discretization size $\Delta
z$ the Courant condition will call for a fixed timestep $\Delta t$
and thus makes the evolution easier to follow numerically. 
    With this time coordinate, the Gowdy metric
[Eq.~(\ref{metric-origgowdy})] becomes
    \begin{equation}
    {ds}^2=t^{-1/2}e^{\lambda/2}(-{dt}^2+{dz}^2)+t\,{dw}^2. 
    \label{metric-tgowdy}
    \end{equation}

With an ingenious parameterization of the metric similar to
Eq.~(\ref{dw-origgowdy}), Gowdy found that the vacuum Einstein
evolution equations could be written in terms of $P$ and $Q$ alone. 
    With the metric in the form of Eq.~(\ref{metric-tgowdy}), these
become
    \begin{eqnarray}
    P_{,tt}+t^{-1}P_t-P_{,zz}-e^{2P}(Q_{,t}^2-Q_{,z}^2)&&=0
\label{pevol} \\[0.5ex]
    Q_{,tt}+t^{-1}Q_t-Q_{,zz}+2(P_{,t}Q_{,t}-P_{,z}Q_{,z})&&=0 ,
\label{qevol}
    \end{eqnarray} 
    with the constraint equations specifying $\lambda$:
    \begin{eqnarray}
    \lambda_{,z}=&&2t(P_{,z}P_{,t}+e^{2P}Q_{,z}Q_{,t})
\label{mc-lambda} \\[0.5ex]
    \lambda_{,t}=&&t[P_{,t}^2+P_{,z}^2+e^{2P}(Q_{,t}^2+Q_{,z}^2)].
    \label{hc-lambda}
    \end{eqnarray}

\subsection{Kasner universe in power-law form}

When the metric parameters $P$ and $Q$ are set to zero, the Gowdy
metric [Eq.~(\ref{metric-tgowdy})] reduces to the simple diagonal
form
    \begin{equation}
    {ds}^2=t^{-1/2}e^{\lambda/2}(-{dt}^2+{dz}^2) +t({dx}^2+{dy}^2).
    \label{metric-kasner}
    \end{equation} 
    This form of the metric represents a homogeneous, anistropically
expanding universe with no gravitational waves and is equivalent to the
form of an axisymmetric Kasner metric \cite{kasner} with a different time
coordinate.  
    A comparison of Eqs.~(\ref{metric-adm}) and
(\ref{metric-kasner}) yields the following analytic solution for the
time evolution of the diagonal components of the field variables
($g_{ij}=K_{ij}=0$ for $i \neq j$) and coordinate conditions:
    \begin{eqnarray}
    g_{11}=g_{22}=t\quad &,& \quad \ g_{33}=t^{-1/2} \quad , 
    \label{gij-kasner}\\[0.5ex]
    K_{11}=K_{22}=-\half t^{1/4}\quad &,& \quad 
                \ K_{33}=\mbox{$\frac{1}{4}$} t^{-5/4} \quad ,
    \label{Kij-kasner}\\[0.5ex]
    \beta^i&=&0 \quad , 
    \end{eqnarray} 
    and\vspace{-1.8\bigskipamount}
    \begin{equation}
    \alpha=\sqrt{g_{33}}=t^{-1/4} \quad , \label{alpha-kasner}
    \end{equation} 
    where we have used Eq.~(\ref{gdot-vacuum}) to determine $K_{ij}$
in terms of $g_{ij,t}$, $\beta^i$, and $\alpha$, and have set
$\lambda$, a constant in this case, to zero.

\subsection{Kasner universe in exponential form}

We can write the metric for the same Kasner universe with a different
time coordinate in an exponential form
    \begin{equation}
    {ds}^2=-e^{t}dt^2+e^{2t/3}dx^2+e^{2t/3}dy^2+e^{-t/3}dz^2,
    \label{metric-expkas}
    \end{equation} 
    which is essentially the Gowdy metric given in
Eq.~(\ref{metric-origgowdy}) with $P=Q=\lambda=0$.  
    A comparison of Eqs.~(\ref{metric-expkas}) and
(\ref{metric-adm}) gives the following analytic solution for
$g_{ii}(t)$ and $K_{ii}(t)$ ($g_{ij}=K_{ij}=0$ for $i\neq j$) and the
coordinate conditions:
    \begin{eqnarray}
    g_{11}=g_{22}=e^{2t/3} \quad &,& \quad
    \ g_{33}=e^{-t/3} \quad , \label{gij-expkas}\\[0.5ex]
    K_{11}=K_{22}=-\mbox{$\frac{1}{3}$}e^{t/6} \quad &,& \quad
    \ K_{33}=\mbox{$\frac{1}{6}$}e^{-5t/6},
    \label{Kij-expkas}\\[0.5ex]
    \beta^i&=&0 \quad ,
    \end{eqnarray} 
    and\vspace{-1.8\bigskipamount}
    \begin{equation}
    \alpha= \sqrt{g} =
    \sqrt{g_{11} \cdot g_{22} \cdot g_{33}}
    =e^{t/2} \quad , \label{alpha-expkas}
    \end{equation} 
    where $g = {\rm det}(g_{ij})$ and we have once again used
Eq.~(\ref{gdot-vacuum}) to determine $K_{ij}$.

\subsection{Polarized waves in an expanding universe}

The Gowdy metric [Eq.~(\ref{metric-tgowdy})] with $Q$ set to zero,
    \begin{equation}
    {ds}^2=t^{-1/2}e^{\lambda/2}(-{dt}^2+{dz}^2)+t(e^P {dx}^2+e^{-P}
{dy}^2),
    \label{metric-polarizedgowdy} 
    \end{equation} 
    represents the spacetime of an expanding universe containing polarized
gravitational waves propagating in the $z$-direction. 
    With this metric, the evolution equations [Eqs.~(\ref {pevol})
and (\ref{qevol})] reduce to a single linear equation for $P$:
    \begin{equation}
    P_{,tt}+t^{-1}P_{,t}-P_{,zz}=0 \label{pevol-polarized}.
    \end{equation} 
    The constraint equations become
    \begin{equation}
    \lambda_{,z}=2tP_{,z}P_{,t} \label{mc-lambda-polarized}
    \end{equation} 
    and
    \begin{equation}
    \lambda_{,t}=t(P_{,t}^2+P_{,z}^2) \label{hc-lambda-polarized}.
    \end{equation}

The general solution to Eq.~(\ref {pevol-polarized}) is a sum of
terms of the form $Z_0(2\pi nt)\cos(2\pi nz)$ and $Z_0(2\pi
nt)\sin(2\pi nz)$, where $n$ is an integer (assuming periodicity of 1
in $z$) and $Z_0$ is a linear combination of the Bessel functions $J_0$
and $Y_0$.
    We have chosen to study the spacetime based on the particular
solution
    \begin{equation}
    P=J_0(2\pi t)\cos(2\pi z). \label{P}
    \end{equation} 
    A comparison of the metrics given in Eqs.~(\ref{metric-adm}) and
(\ref{metric-polarizedgowdy}) then yields the following exact
solution for the time evolution of the diagonal components of the
field variables ($g_{ij}=K_{ij}=0$ for $i \neq j$) and coordinate
conditions:
    \begin{equation}
    g_{11}=te^P,\ g_{22}=te^{-P},\ g_{33}=t^{-1/2}e^{\lambda/2},
    \label{g_ij-polarizedgowdy}
    \end{equation}
    \begin{eqnarray}
    K_{11}&=&-\mbox{$\frac{1}{2}$}t^{1/4}
                        e^{-\lambda/4}e^P(1+tP_{,t}),
    \nonumber \\[0.5ex]
    K_{22}&=&-\mbox{$\frac{1}{2}$}t^{1/4}
                        e^{-\lambda/4}e^{-P}(1-tP_{,t}),
    \label{K_ij-polarizedgowdy} \\[0.5ex]
    K_{33}&=&\mbox{$\frac{1}{4}$}t^{-1/4}
                        e^{\lambda/4}(t^{-1}-\lambda_{,t}),
    \nonumber
    \end{eqnarray}
    \begin{equation}
    \beta^i=0,
    \end{equation} 
    and
    \begin{equation}
    \alpha=\sqrt{g_{33}}=t^{-1/4}e^{\lambda/4},
\label{alpha-polarizedgowdy}
    \end{equation} 
    where we have again used Eq.~(\ref{gdot-vacuum}) to determine
$K_{ij}$.
    This set of equations is completed by an expression for
$\lambda$, which can be derived by using Eq.~(\ref{P}) in
conjunction with Eqs.~(\ref{mc-lambda-polarized}) and
(\ref{hc-lambda-polarized}):
    \begin{eqnarray}
    \lambda&=&-2\pi t J_{0}(2\pi t) J_{1}(2\pi t) \cos^{2}(2\pi z) 
    \\[0.7ex]
    & & \mbox{} + 2\pi^{2}t^{2}
    \bigl[J_{0}^{2}(2\pi t) + J_{1}^{2}(2\pi t)\bigr]
           \nonumber \\[0.5ex]
    & & \mbox{} - \half
    \bigl\{ (2\pi )^{2}\bigl[J_{0}^{2}(2\pi )
                             +J_{1}^{2}(2\pi )\bigr]-
    2\pi J_{0}(2\pi ) J_{1}(2\pi )\bigr\} \label{lambda}.
    \nonumber
    \end{eqnarray}

\section{Numerical Relativity Simulations} \label{sims}

We have used the classes of Gowdy spacetimes discussed in 
Sec.~\ref{gowdy} as testbeds for numerical relativity
simulations using the deloused leapfrog, staggered leapfrog, and
Crank-Nicholson time integration techniques discussed in 
Sec.~\ref{instability}.
    We have carried out these simulations using two types of codes,
toy codes (which solve the Einstein equations Eqs.~(\ref{gdot-ode})
and (\ref{kdot-ode}) for spatially homogeneous spacetimes) and the
ADM code developed by the Binary Black Hole (BBH) Grand Challenge
Alliance \cite{alliance,alliance_prl} (which solves the full vacuum
Einstein equations Eqs.~(\ref{gdot-vacuum}) and (\ref{kdot-vacuum})
in three spatial dimensions).  
    In this section, we present the results of these runs plus
efficiency analyses that compare the numerical cost effectiveness of
the various time integration schemes.
 
\subsection{Toy code simulations} \label{toy}

The vacuum Einstein equations for spatially homogeneous metrics
reduce to the set of ordinary differential equations 
Eqs.~(\ref{gdot-ode}) and (\ref{kdot-ode}).   
    These are similar in form to the set of equations Eqs.\
(\ref{newt}), with the three-metric $g_{ij}$ replacing $\vec r$ and
the extrinsic curvature $K_{ij}$ replacing $\vec v$.  
    Note that, as in the case of the full Einstein equations [Eqs.\
(\ref{gdot-vacuum}) and (\ref{kdot-vacuum})], the system of Eqs.\
(\ref{gdot-ode}) and Eqs.\ (\ref{kdot-ode}) generally contains
``velocity-dependent forces,'' and thus 3-level leapfrog integrations
of these equations will be inherently unstable.
    We have constructed toy codes to solve these equations using
each of the integration methods discussed in Sec.\ \ref{instability}. 
    The 3-level leapfrog toy code is based on the discretization in
Eqs.\ (\ref{discrete-newt}), with the delousing module based on the
Steps (1) - (5) outlined in Sec.\ \ref{deloused}.  
    The staggered leapfrog toy code is based on the discretization in
Eqs.\ (\ref{staggered}); in this method we use the extrapolation
$K^{n+1}_{ij}=(3/2)K^{n+1/2}_{ij}-(1/2)K^{n-1/2}_{ij}$ to obtain $K_{ij}$ on
the full integer levels \cite{anninos-holes}.  
    Although this extrapolation is accurate to second-order in
$\Delta t$, it is not time symmetric.
    Finally, the Crank-Nicholson toy code is based on the
discretization in Eqs.\ (\ref{cn-init}) - (\ref{cn2}).

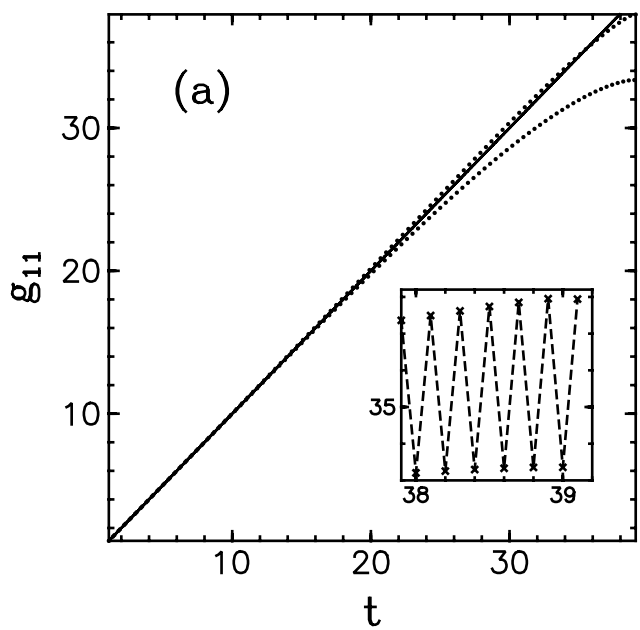
\begin{figure}
    \caption[3-level leapfrog, Kasner]{The results of the unstable
3-level leapfrog toy code integration of a Kasner universe in
power-law form are presented here. 
    The numerical (dots, with every other pair of 
even and odd timestep values plotted) and analytical (solid lines) solutions
for $g_{11}(t)$ and $K_{11}(t)$ are given in panels a and b, respectively.
    The inset in panel a is an enlargement of the numerical solution
for $g_{11}(t)$ in the range $37.9<t<39.2$, in which all data points
have been plotted and connected with dashed lines to emphasize the
large alternate timestep oscillations of this unstable solution.  
    The numerical accuracy measures $H_{norm}(t)$ and $e_{g}(t)$ are
shown in panels c and d, respectively; for the sake of clarity, only
every other pair of even and odd timestep values has been plotted.}
\label{plkas-unstable} 
\end{figure}

A toy code simulation of the Kasner universe in power-law form
from Sec.~V.{}A was the initial testbed numerical relativity problem
to which we applied the deloused leapfrog technique.
    To demonstrate the need for the delousing modification of the
standard 3-level leapfrog technique, we first carried out a run
with the standard 3-level leapfrog technique itself.
    Eqs.~(\ref{gij-kasner})-(\ref{alpha-kasner}) provide both the
initial conditions for this simulation (begun at $t=1$ and run with a
timestep $\Delta t=0.1$) and the means to measure its accuracy via a
comparison between the analytically and numerically determined values
of the field variables.  
    Specifically, we use
    \begin{equation}
    e_g=\Bigl[\sum_{i=1}^{3}
\bigl(\frac{g_{ii}}{g^{an}_{ii}}-1\bigr)^2\Bigr]^
    {1/2} \label{eg}
    \end{equation} 
    as a measure of a simulation's accuracy.  
    Here the analytic values of the diagonal components
$g_{ii}$ [given in this case by Eq.\
(\ref{gij-kasner})] are denoted as $g^{an}_{ii}$.
    Another quantity useful in determining the quality of the
integration is the size of the normalized residual of the vacuum
Hamiltonian constraint [Eq.~(\ref {hc-vacuum})]
    \begin{equation}
    H_{norm}=\frac{\mid\! R+K^2-K_{ij}K^{ij}\!\mid}
    {\mid\! R\!\mid+K^{j}_{i}K^{i}_{j}}. \label{hnorm}
    \end{equation} 
$R^{i}_{i}=R=0$ for
the spatially homogeneous Gowdy spacetimes.
    The results of the 3-level leapfrog simulation are presented in
Fig.~\ref{plkas-unstable}.
    Panels a and b of this figure display both the analytical (solid
lines) and numerical (dots) solutions for the field variable
components $g_{11}(t)$ and $K_{11}(t)$, respectively; panels c and d
show the accuracy measures $H_{norm}(t)$ and $e_{g}(t)$,
respectively.
    The separate even and odd timestep numerical solutions,
characteristic of the leapfrog instability described in Sec.\ 
\ref{instability} (cf.\ Eq.\ (\ref{z-x-y})), are clearly visible in
all four panels of Fig.\ \ref{plkas-unstable} as two dotted branches
representing the numerical solution.
    The fact that these two dotted branches are indeed alternate
timestep oscillations of the numerical solution is evident in the
inset of Fig.~\ref{plkas-unstable}a, which is an enlargement of the
numerical solution of $g_{11}$ for $37.9<t<39.2$.  
    In this inset every value of $g_{11}$ has been plotted with an
``x'' and consecutive values have been connected by dashed lines.

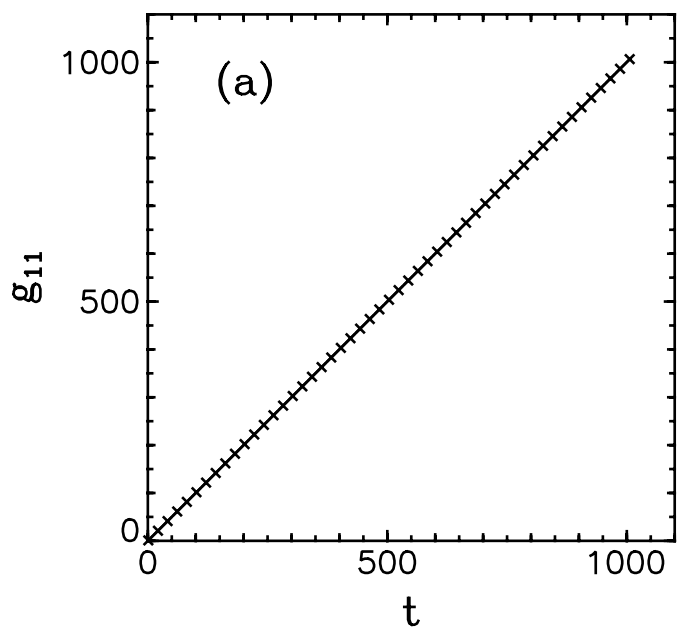
\begin{figure}
    \caption[Deloused leapfrog, Kasner]{This figure depicts the stable
deloused leapfrog, toy code integration begun with the same
initial conditions as the unstable 3-level leapfrog simulation
presented in Fig.~\ref{plkas-unstable}. 
    The numerical (``x''s, with every 201st point plotted) and
analytical (solid line) solutions for $g_{11}(t)$ and $K_{11}(t)$ are
given in panels a and b, respectively.  
    The accuracy measures $\log(H_{norm}(t))$ and $e_{g}(t)$ are
shown in the lower panels c and d, respectively.}
\label{plkas-stable} 
\end{figure}

A toy code integration starting with the same initial conditions
as the 3-level leapfrog run depicted in Fig.~\ref{plkas-unstable} was
also performed with the deloused leapfrog technique.
    The trigger chosen to initiate the delousing steps in this run was
a change in the sign of the slope of $H_{norm}(t)$; such a sign
change is indicative of the alternate timestep oscillations discussed
in the preceeding paragraph.
    The results of this integration are given in
Fig.~\ref{plkas-stable}.
    Panels a and b of this figure display the analytical (solid lines)
and numerical (``x''s) solutions for $g_{11}(t)$ and $K_{11}(t)$,
respectively; the solid lines in panels c and d represent $\log
(H_{norm}(t))$ and $e_{g}(t)$, respectively.  
    The regular behavior of the evolved quantities shown in  Fig.\
\ref{plkas-stable} demonstrates that the delousing steps successfully
removed the parasitic mode from the solutions for the field
variables, producing an evolution that was stable for the duration of
the integration. 
    We ran the deloused code a factor of 25 times longer than the
duration of the catastrophically unstable 3-level leapfrog
simulation.

    Note that in general it is possible to evolve components of
$g_{ij}$ or $K_{ij}$ using the constraint equations 
[Eqs.~(\ref{hc-vacuum}-\ref{mc-vacuum})] in place of one or more
of the evolution equations [Eqs.~(\ref{gdot-vacuum}-\ref{kdot-vacuum})].  
    For integrations of spatially homogeneous spacetimes,
only the Hamiltonian constraint is meaningful in this context.
    In order to investigate the stability of such constrained evolutions, we
performed 3-level leapfrog toy code simulations of the Kasner universe
in exponential form in which we replaced the evolution equation for $K_{33}$
[Eq.~(\ref{kdot-ode}) with $i=j=3$] with Eq.~(\ref{hc-vacuum}).
    Thus $K_{33}$ was calculated in terms of the evolved quantities
$K_{11}$ and $K_{22}$ by imposing
the Hamiltonian constraint.
    These constrained runs still suffered from the instability under discussion.
    However, the times at which the simulations became catastrophically unstable
were almost two and a half times longer than in the corresponding
unconstrained runs.

We also carried out integrations of this model using the staggered
leapfrog and Crank-Nicholson techniques.  
    These runs were all stable, as expected.

\subsubsection*{Efficiency analysis of toy code runs}

The efficiency of a stable integration technique is also an important
factor to consider in evaluating numerical methods.  
    Here, we consider the efficiency of a technique to be the
accuracy level it maintains for a particular numerical cost.  
    Since the evaluation of the right-hand sides of the discretized
equations [e.g., Eqs.\ (\ref{discrete-newt}), (\ref{staggered}), and
(\ref{cn2}) for the set of equations Eqs.\ (\ref{newt})] is generally the most
expensive operation in terms of CPU time, we define the cost of an
integration to be the number of times the right-hand sides are
computed.

The results of our efficiency comparison for the toy code
simulations of the Kasner universe in power-law form are displayed in
Fig.~\ref{eff-plkas}.  
    For this comparison, we ran simulations with each of the three
stable integration methods; in these simulations the initial
conditions and evolution duration were identical but the constant
timestep used during the simulations $\Delta t$ was varied from run
to run.  
    We used the value of $e_g$ at the end of the
simulation as the accuracy measure.
    The left panel of Fig.\ \ref{eff-plkas} gives the final values of 
$\log (e_{g})$ as a function of $\log(\Delta t)$ and demonstrates 
that all three techniques are second-order
accurate (i.e.{}, the slope of $\log (e_{g})$ versus $\log(\Delta t)$ 
for each method is $\sim 2$). 
    The right panel of Fig.\ \ref{eff-plkas} gives the final values
of $\log (e_{g})$ as a function of the numerical cost
(measured by the number of times both $dg_{ij}/dt$ and $dK_{ij}/dt$
are computed) and provides the most informative picture of the
efficiency of the different techniques.
    The average number of iterations per timestep for the
Crank-Nicholson runs ranged between two and three [for a convergence
criterion of $1.0\times 10^{-8}$; see Sec.\ II]\@.  
    The total number of delousing steps performed during the deloused
leapfrog runs ranged from 153 to 181.  
    Note that each delousing step adds eight calls to the cost of the
integration since it requires four calls to the RK2 routine (see
Sec.\ III), which in turn computes $dg_{ij}/dt$ and $dK_{ij}/dt$
twice. 
    Panel b shows that, for simulations of this simple,
spatially invariant spacetime, Crank-Nicholson is the most efficient
of the three integrators. 
    We have used least-squares analysis to fit the best straight lines
to the data points shown in Fig.\ \ref{eff-plkas}b.
    This analysis led to the following relationship for $e_{g}({\rm cost})$:
    \begin{equation}
    e_g=10^{b_g}{\rm cost}^{m_g}. \label{egfit}
    \end{equation}
    The values of the parameters $b_g$ and $m_g$ for these fits are given 
for each integrator in Table 1.

We have also done an efficiency comparison of these three integration
techniques for toy code simulations of the Kasner universe in
exponential form; the results are shown in Fig.~\ref{eff-expkas}.
    The outcome of the efficiency tests for these simulations is
quite different from that based on the Kasner universe in power-law
form of Fig.~\ref{eff-plkas}. 
    Fig.~\ref{eff-expkas}b shows that for
relatively low to moderate cost and accuracy demands, the staggered
leapfrog method is the most cost effective technique in this case;
however, the deloused leapfrog method is more efficient when high
accuracy levels are required.
    The higher average number of iterations per timestep required by
these Crank-Nicholson runs, which ranged from three for $\Delta
t=0.0016$ to seven for $\Delta t=0.1$, may account for the reversal
in its relative cost effectiveness from the simulations of the
power-law form (Fig.~\ref{eff-plkas}).  
    The number of delousing steps taken during the deloused leapfrog
runs ranged from 1285 for $\Delta t=0.0016$ to 422 for $\Delta
t=0.1$.  
    Thus the deloused leapfrog method had to work harder to maintain
stable integrations of this universe in exponential form than in
power-law form.
    We have again used least squares analysis to fit straight lines
to the data in Fig.~\ref{eff-expkas}b and 
produce a relation of the form of Eq.\ (\ref{egfit});
this relation is parameterized by the values of $b_g$ and $m_g$ 
given in Table 2.

\subsection{ADM code simulations} \label{adm}

Our study of the deloused leapfrog method stems from our search for
an efficient technique capable of performing numerically stable
simulations of the orbital dynamics of binary neutron stars.
    Because such simulations require the solution of the full
Einstein equations, we wanted to test the deloused leapfrog
integrator in conjunction with the code we plan to use to do these
simulations, the ADM code developed by the BBH Alliance
\cite{alliance,alliance_prl}.
    This second-order accurate code currently solves the vacuum
Einstein equations on a Cartesian grid and provides the user the
choice of utilizing either the standard 3-level leapfrog or
Crank-Nicholson integration techniques.
    We have added the capability of using the deloused leapfrog
integrator to the BBH Alliance's ADM code and have used it to perform
simulations of a Kasner (homogeneous Gowdy) expanding spacetime with
the power-law coordinate condition of Sec.~V.{}A and of the expanding
Gowdy spacetime with polarized gravitational waves of Sec.~V.C\@.
    Of course this code, like the toy code, was ignorant of
Gowdy's ingenuity which, through parameterizations like $g_{11}=
t\,e^P$, can reduce some of the Einstein equations to linear
equations.
    The Einstein equations are coded in terms of the $g_{ij}$ and
$K_{ij}$ as shown in Eqs.~(\ref{gdot-vacuum}) and (\ref{kdot-vacuum});
the chosen coordinate conditions were
$\beta^i = 0$ and $\alpha=\sqrt{g_{33}}$.
    They involve not only the polynomial nonlinearities manifest in
these equations, but also the nonlinearities implied through
$\alpha$ and through the inverse metric when indices are
raised or covariant derivatives or curvatures are computed.

The preliminary testbed used in our ADM code runs was a simulation
identical to the toy code runs of the Kasner power-law metric.
    The development of the instability in the ADM code's 3-level
leapfrog run replicated its development in the toy code run. 
    The ADM code's deloused leapfrog run successfully removed this
instability in the same manner as in the toy code run, with the
delousing steps triggered at the same temporal intervals in both
simulations.

To test the behavior of the deloused leapfrog method for partial
differential equations with spatially varying terms, we carried out
simulations of the polarized Gowdy spacetime of Sec.~V.B\@.
Equations~(\ref{P})-(\ref{hc-lambda}) yield both initial
conditions for simulations of this spacetime and exact solutions with
which to compare the results of such simulations.

Our ADM code polarized Gowdy simulations began at $t=1$ and were run
with periodic boundary conditions over the interval $-\half\leq z \leq \half$
and a grid spacing $\Delta z=1/62$.
    Because the vacuum Einstein equations are partial differential
equations, the size of the timestep that can be taken in the
integration is restricted by the Courant condition \cite{nr}, which
ensures that information cannot propagate across more than a single
grid zone in one timestep.  
    For the polarized Gowdy metric of 
Eq.~(\ref{metric-polarizedgowdy}), this condition is equivalent to
enforcing $\Delta t=C\, \Delta z$, where $\Delta z$ is the (uniform)
grid spacing and the Courant factor $C<1$.
    In the runs presented here, we chose $C=0.3$.
    The initial ADM code simulation was performed with the 3-level
leapfrog integrator and, as expected, was unstable.  
    The results of the integration are presented in 
Figs.~\ref{gowdy-unstable1} and \ref{gowdy-unstable2}.
    The evolution of the value of $g_{11}$ at the center of the grid
is shown in panel a of Fig.~\ref{gowdy-unstable1}; panels b, c, and d
of this figure display, respectively, $\overline{H}_{norm}(t)$,
$\overline{e}_g(t)$, and $\overline{e}_t(t)$.  
    Here bars denote (spatial) averages over the grid. 
    $\overline{H}_{norm}(t)$ is computed by dividing the 
spatial average of the numerator of
Eq.~(\ref{hnorm}) by the spatial average of the denominator of
Eq.~(\ref{hnorm}).
    The additional measure of the accuracy of the simulation, $e_t$, is
defined by
    \begin{equation}
    e_{t}(x,y,t)=t^{-1}\mid\!
t-[g_{11}(x,y,x)\,g_{22}(x,y,x)]^{1/2}\!\mid.
    \label{et}
    \end{equation} 
    The usefulness of $e_t$ as an error estimate arises because,
according to the analytic solution of 
Eq.~(\ref{g_ij-polarizedgowdy}), $g_{11}g_{22}=t^2$.  

    The separate even and odd timestep solutions, indicative of the
instability, can clearly be seen in the plots of
$\overline{H}_{norm}(t)$, $\overline{e}_g(t)$, and
$\overline{e}_t(t)$; however, the separate solutions are not yet
visible in the plot of $g_{11}$ at $t=5.8$.
    As shown in Fig.~\ref{gowdy-unstable2}, they do appear in
$g_{11}$ at the grid center later in the evolution, as the
instability begins to overwhelm the computation.

We then evolved the same polarized Gowdy initial data with the ADM
code using the deloused leapfrog integrator.  
    Because $\overline{H}_{norm}(t)$ and $\overline{e}_g(t)$ oscillated in our 
simulations of this spacetime, they were not used as a basis for the
trigger that initiated the delousing steps.  [These fluctuations are
due to the complex oscillatory nature of $g_{ij}$ and $K_{ij}$ in this
spacetime and are {\it not} related to the alternate timestep oscillations 
caused by the instability; in fact, such fluctuations are also present in the
Crank-Nicholson simulations of this spacetime.]
    Instead, because $\overline{e}_t$ behaved monotonically (see
Fig.~\ref{gowdy-stable}d), a change in the sign of its temporal 
slope was used as the delousing trigger.  
    As can be seen in Fig.~\ref{gowdy-stable}, the removal of the
parasitic mode during the delousing steps taken in this simulation
eliminated the presence of large alternate timestep oscillations and
allowed for a stable integration. 

\subsubsection*{Efficiency analysis of ADM code runs}

We have also used simulations of this polarized Gowdy spacetime to
evaluate the efficiency of the deloused leapfrog algorithm.
    However, in this case its performance could only be compared with that of
the Crank-Nicholson technique, as the staggered leapfrog method has
not been implemented in the ADM code. 
    (The reason for this is that the memory requirements of the
staggered leapfrog method would exceed those of the other two methods
if adaptive mesh refinement were to be used.) In this efficiency
comparison, $\Delta z$ and $\Delta t$ are reduced in tandem from run
to run, with $C$ held constant at 0.3.
     The value of $\overline{e}_{t}$ at the end of the simulations
was used as the measure of accuracy upon which to base the efficiency analysis
in this case.

The results of this analysis are displayed in 
Fig.~\ref{eff-polarizedgowdy}.
    The left panel of this figure gives the final value of
$\overline{e}_{t}$ as a function of the grid spacing $\Delta z$,
which was varied from $1/126$ to $1/62$ to $1/30$.
    Straight lines fit throught these data points have slopes $\sim
2$; this demonstrates the second-order accuracy of the deloused
leapfrog and Crank-Nicholson methods.  
    The right panel contains a plot of $\overline{e}_{t}$ versus cost
and indicates that the deloused leapfrog integrations of this
spacetime were about five to eight times more efficient than those
carried out with the Crank-Nicholson technique.  
    The average number of iterations per timestep for the
Crank-Nicholson runs ranged from five for $\Delta z=1/126$
to eight for $\Delta z=1/30$.  
    The total number of delousing steps taken was relatively constant
in the three deloused leapfrog runs (six for $\Delta z=1/126$; five
for $\Delta z=1/62$; and six for $\Delta z=1/30$).  
    The least squares straight line fits to the data in 
Fig.~\ref{eff-polarizedgowdy}b can be transformed, in this case, to
relations of the form 
    \begin{equation}
    e_t=10^{b_t}{\rm cost}^{m_t}; \label{etfit}
    \end{equation} 
    the parameters $b_t$ and $m_t$ determined by these fits are given
in Table 3.

\section{Conclusions} \label{conclusions}

    The purpose of this paper is to alert the community to the existence
of the instability inherent in standard 3-level leapfrog integrations
of Einstein's equations and to demonstrate that the proposed delousing
modification to the standard algorithm can efficiently cure this
instability.
    To this end, we have used three classes of testbed solutions.
    In Sec.~\ref{deloused}, we show calculations of highly relativistic
circular geodesic orbits calculated in Cartesian coordinates.
    In Sec.~\ref{sims}, we show evolutions of homogeneous expanding
cosmologies and polarized gravitational waves in an expanding
Gowdy spacetime. 
    In all of these cases, the deloused leapfrog algorithm removed
parasitic modes from the numerical solution of the Einstein equations 
that were excited to instability in traditional 3-level leapfrog 
simulations of these spacetimes, and thus allowed for their stable evolution.

The numerical efficiency (i.{}e.{}, the accuracy level maintained for
a particular numerical cost) of the deloused leapfrog integrator was
compared to the efficiencies of two other stable integration
methods, staggered leapfrog and Crank-Nicholson.  
    We have defined numerical cost as the number of times the right
hand sides of both Einstein evolution equations (i.{}e.{}, $g_{ij,t}$
and $K_{ij,t}$) are computed during a simulation.  
    Thus cost in this case is a measure of a simulation's CPU
expense. 
    Note that all of the simulations presented in this paper were
carried out with constant timesteps.

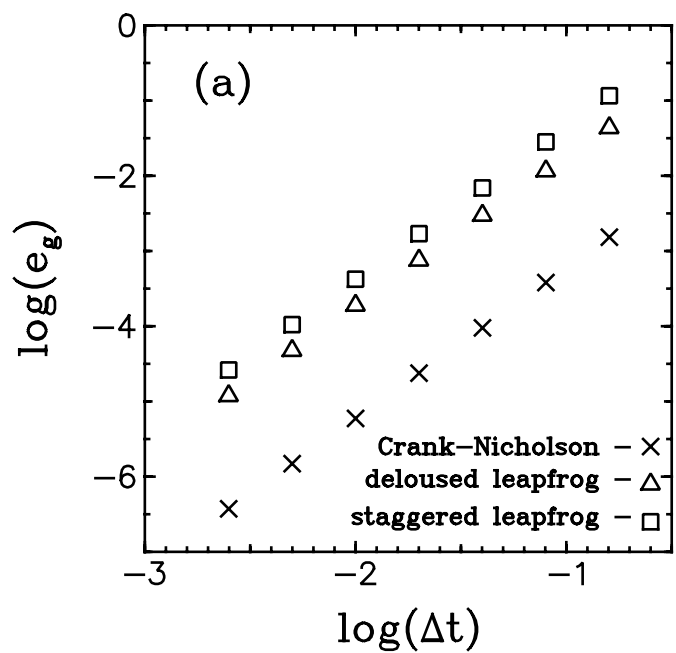
\begin{figure}
    \caption[Efficiency in powerlaw Kasner test.]{The results of the
efficiency analysis of toy code simulations of a Kasner universe in
power-law form, represented by the metric given in
Eq.~(\ref{metric-kasner}), carried out with different integration
techniques are shown here. 
    The final values of $\log(e_g)$ are plotted
versus $\log(\Delta t)$ in the left panel and versus $\log({\rm
cost})$ in the right panel.
    Values from Crank-Nicholson runs are marked with ``x''s, those
from deloused leapfrog runs are marked with triangles, and those from
staggered leapfrog runs are marked with squares. 
    The cost is defined as the number of times the right-hand sides
of the discretized equations are evaluated.  
    The slopes of straight lines fit through data in (a) are
$\sim 2$, indicating that the numerical techniques are all accurate
to second-order in $\Delta t$.} 
\label{eff-plkas} 
\end{figure}

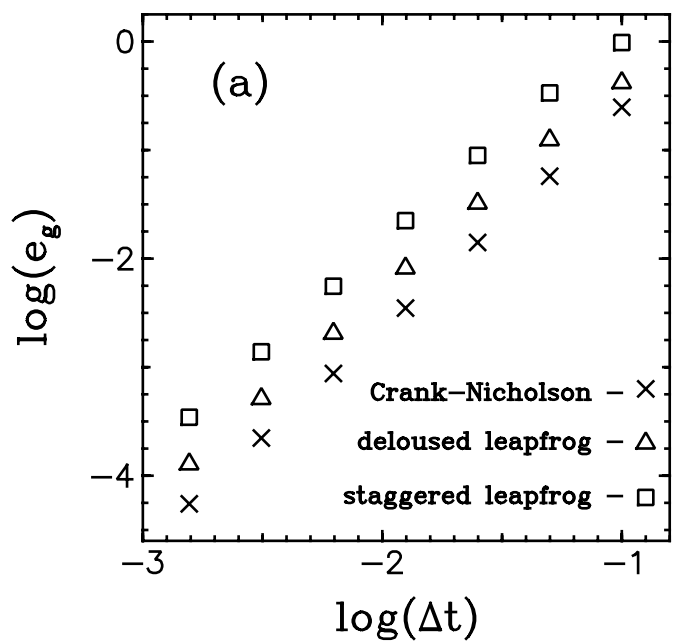
\begin{figure}
    \caption[Efficiency in exponential Kasner test.]{The same
quantities and notation as in Fig.~\ref{eff-plkas}, but for toy code
simulations of a Kasner universe in exponential form, described by
the metric of Eq.~(\ref{metric-expkas}).} 
    \label{eff-expkas} 
    \end{figure}

The first testbed for this efficiency analysis was a toy code
(which solves the spatially invariant Einstein equations 
[Eqs.~(\ref{gdot-ode}) and ((\ref{kdot-ode})]) simulation of a 
spatially homogeneous Gowdy spacetime yielding a Kasner universe.
    With the coordinate condition choice $\alpha=\sqrt{g_{33}}$ this
gives a power-law analytic solution for the metric components.
    For this simple problem, Crank-Nicholson was the most efficient
of the integrators (see Fig.\ \ref{eff-plkas}).  
    The results were different, however, when the testbed was changed
to a toy code simulation of the same Kasner universe with a different
time coordinate choice $\alpha=\sqrt{g}= \sqrt{g_{11} \cdot g_{22}
\cdot g_{33}}$, yielding in the analytic solution an {\it
exponential} form for the metric components.
    In that case, the staggered and deloused leapfrog techniques were
more cost effective (see Fig.~\ref{eff-expkas}), as the rapid
evolution of the spacetime caused the iterative Crank-Nicholson
technique to require a larger number of iterations per timestep.

The final testbed used in our efficiency analysis was a simulation of
an expanding Gowdy spacetime containing polarized gravitational
waves.  
    These simulations required the solution of the complete vacuum
Einstein equations and were carried out with the BBH
Alliance's ADM code \cite{alliance,alliance_prl}, in which the standard
(unstable) 3-level leapfrog and the (stable) Crank-Nicholson integration
methods had previously been implemented.  We modified this code to allow
the use of the deloused leapfrog scheme.  
    Because the staggered leapfrog method has not been implemented in
the ADM code, the efficiencies of only the Crank-Nicholson and
deloused leapfrog integrators were evaluated in this case.
    The deloused leapfrog integrations of this spacetime were five to
eight times more cost effective than the Crank-Nicholson runs (see
Fig.\ \ref{eff-polarizedgowdy}).

    Thus, the results of our testbed simulations indicate that the deloused
leapfrog algorithm is an effective and efficient cure for the 3-level
leapfrog instability.  
     Further evaluation of this algorithm, via its
use in more complex problems in numerical relativity, such as a
contracting Gowdy universe containing unpolarized gravitational waves,
and other fields, would serve to confirm the robustness of the method
and provide insight into its cost effectiveness in different numerical
scenarios.

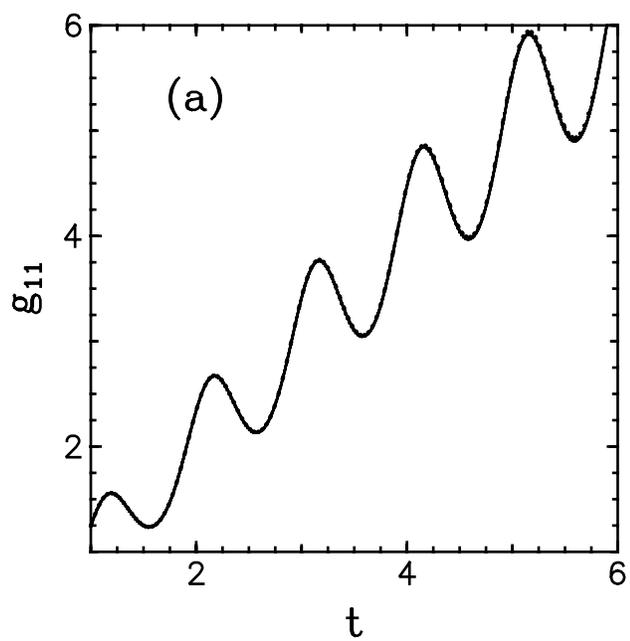
\begin{figure}
    \caption[3-level leapfrog, Gowdy polarized.]{The results of the
unstable 3-level leapfrog ADM code integration of the expanding
universe containing polarized gravitational waves are presented
here.  
    The numerical (dots, with every fourth pair of even and odd
timestep data points plotted)
 and analytical (solid line) solutions for $g_{11}(t)$ at the grid
center are given in panel a.  
    The numerical accuracy measures $\overline{H}_{norm}$,
$\overline{e}_g$, and $\overline{e}_t$ are plotted as functions of
time in panels b, c, and d, respectively; again, only every fourth
pair of even and odd timestep values has been plotted.} 
\label{gowdy-unstable1} 
\end{figure}

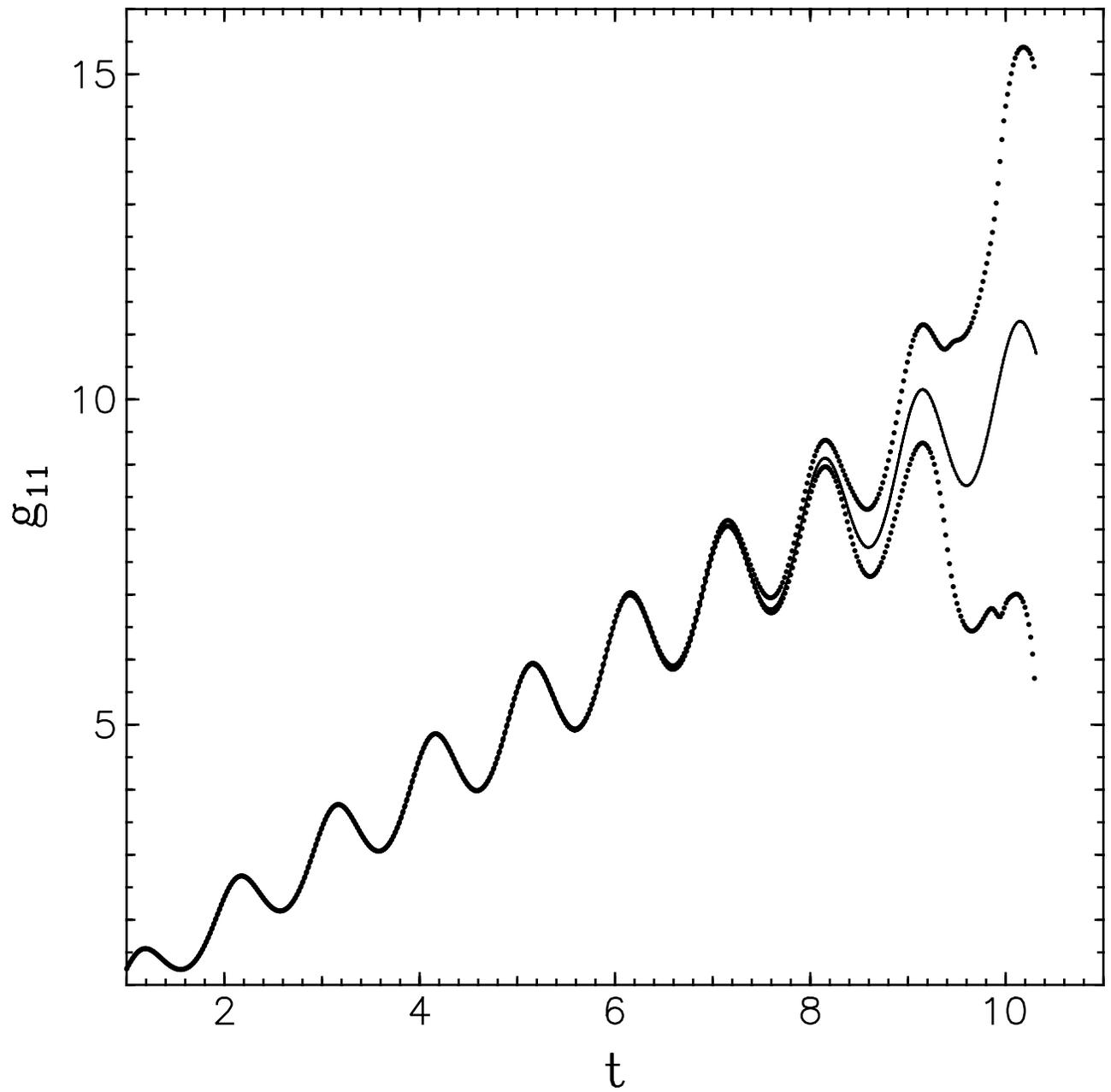
\begin{figure}
    \caption[3-level leapfrog, Gowdy polarized, detail.]{The extended
evolution of the metric component $g_{11}$, at the grid center, from
the unstable 3-level leapfrog simulation presented in Fig.\
\ref{gowdy-unstable1} is shown here.  
    The notation is the same as that of panel a in Fig.\
\ref{gowdy-unstable1} (except that every other pair of even and odd
timestep values is plotted), but the duration of the evolution has
been extended to exhibit the growth of the parasitic mode, as
evidenced by the appearance of the even and odd timestep branches of
the numerical solution.} 
\label{gowdy-unstable2} 
\end{figure}

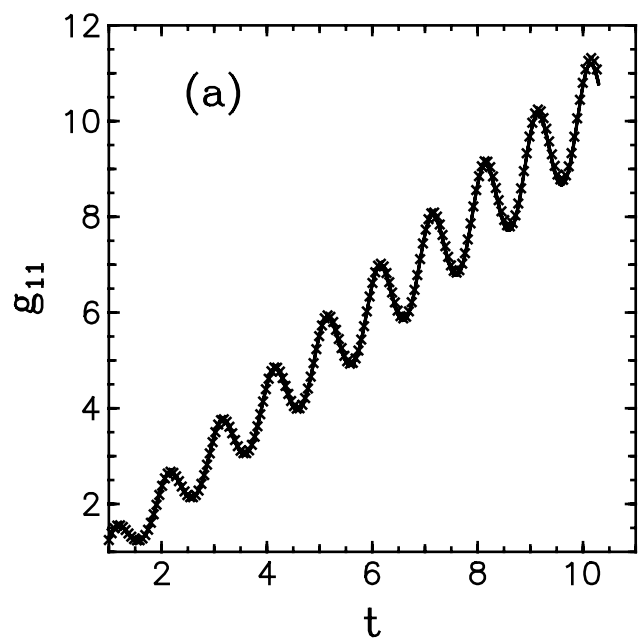
\begin{figure} 
    \caption[Deloused leapfrog, Gowdy polarized.]{This figure depicts
the stable deloused leapfrog, ADM code integration begun with the
same initial conditions as the unstable 3-level leapfrog simulations
shown in Figs.\ \ref{gowdy-unstable1} and \ref{gowdy-unstable2}.  
    The numerical (``x''s, with every eleventh point plotted) and
analytical (solid line) solutions for $g_{11}(t)$ at the center of
the grid are given in panel a.  
    Panels b, c, and d present the evolutions of the accuracy
measures $\overline{H}_{norm}$, $\overline{e}_g$, and
$\overline{e}_t$, respectively.} 
\label{gowdy-stable} 
\end{figure}

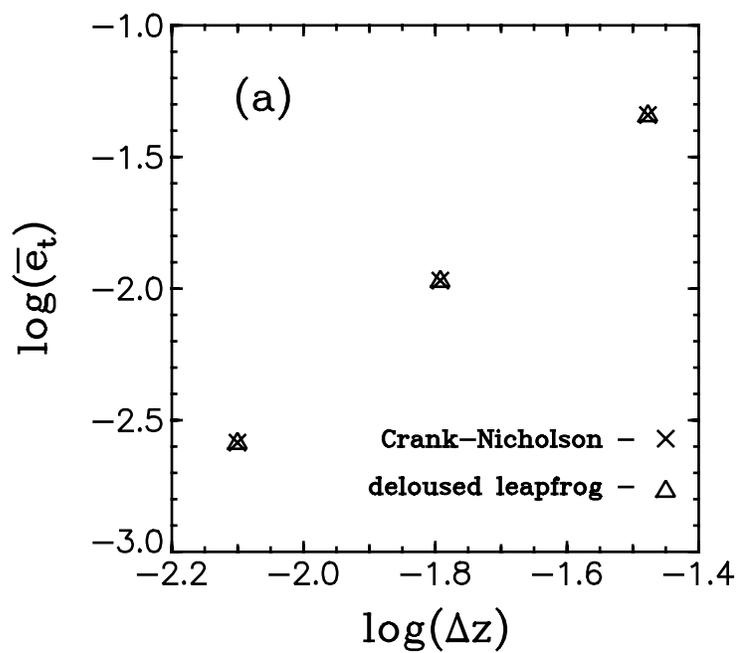
\begin{figure}
    \caption[Efficiency comparisons, polarized Gowdy.]{The results of
the efficiency analysis of ADM code simulations of an
expanding universe containing polarized gravitational waves,
represented by the metric of Eq.\ (\ref{metric-polarizedgowdy}), are
shown here.  
    The final values of $\log(\overline{e}_t)$ are plotted as a
function of the logarithm of the grid spacing $\Delta z$ ($\Delta
t=0.3\,\Delta z$) in the left panel and as a function of the
logarithim of the numerical cost in the right panel.  
    Triangles mark the data points from deloused leapfrog runs;
``x''s mark those from Crank-Nicholson runs.  Straight lines fit
through data points in the left panel have slope $\sim 2$, indicating
that the numerical techniques are both accurate to second-order in
$\Delta t$.} 
\label{eff-polarizedgowdy} 
\end{figure}

    One aspect of the deloused leapfrog algorithm that has the potential
to alter the conclusions of such efficiency analyses is the choice of
delousing trigger.  
    Based on our experience, we suspect that choice of a trigger 
which initiates an {\em excessive} number of delousing steps will degrade
the accuracy of a simulation to some degree.
    If this is the case, a decrease in the average interval
between delousing steps would not only increase the
cost of the run, but would also decrease its numerical accuracy somewhat.
    On the other hand, an increase in the delousing interval
would allow the parasitic mode in the numerical solution
to grow to higher levels.
    For example, had a change in the sign of the temporal slope of 
$g_{11}(t)$ been used as the delousing trigger in the deloused leapfrog
simulation of the polarized Gowdy spacetime, the parasitic mode would
likely have grown to a greater extent between delousing intervals, as
its presence became sizeable in $g_{11}(t)$ rather late in the
evolution (see Figs.\ \ref{gowdy-unstable1} and
\ref{gowdy-unstable2}).
    Thus the choice of delousing trigger may involve a trade-off between
the loss of some degree of accuracy introduced into the computation
via the delousing steps
and the degree to which the parasitic mode is permitted to grow
between these steps.  

In conclusion, we have demonstrated that the deloused leapfrog
algorithm permits the stable numerical evolution of simple vacuum
spacetimes.  
    In addition, our results suggest that deloused leapfrog may be a
better integration technique than Crank-Nicholson to employ in
complex numerical relativity simulations, as this new algorithm was
more cost effective than the Crank-Nicholson method in our simulation
of a spatially varying spacetime.

\acknowledgments
    We thank Matt Choptuik,
Alex Dragt, Mijan Huq, Scott Klasky, Steve McMillan, and
Conrad Schiff for interesting and helpful discussions.  
    We are grateful to the Binary Black Hole Alliance (NSF ASC/PHY
938152-ARPA Supplemented, R. Matzner PI) for making their ADM code
available to us, and and to Mijan Huq and Scott Klasky for helping us to
learn how to run the code.
    We also thank the anonymous referee for thoughtful comments
that resulted in improvements in this manuscript.

    This work was supported in part by NSF grants PHY 9208914 and PHY
9722109 at Drexel, and PHY 9700672 at the University of Maryland.  The
numerical simulations using the ADM code were run at the Northeast
Parallel Architectures Center (NPAC) at Syracuse University.


%
%

%
\begin{table}
\caption[Fits1.]{Parameters for fits to accuracy versus cost data in
Fig.~\ref{eff-plkas}}
    \label{table1}
    \begin{tabular}{c c c}
Method & $b_g$ & $m_g$ \\
\hline
deloused leapfrog & 6.6 & -2.0 \\
Crank-Nicholson & 6.8 & -2.2 \\
staggered leapfrog & 6.7 & -2.0 
\end{tabular}
    \end{table}
    \begin{table}
\caption[Fits2.]{Parameters for fits to accuracy versus cost data in
Fig.~\ref{eff-expkas}}
    \label{table2}
\begin{tabular}{c c c}
Method & $b_g$ & $m_g$ \\
\hline
deloused leapfrog & 11 & -3.1 \\
Crank-Nicholson & 9.0 & -2.6 \\
staggered leapfrog & 5.8 & -1.9 
    \end{tabular}
    \end{table}
    \begin{table}
\caption[Fits3.]{Parameters for fits to accuracy versus cost data in
Fig.~\ref{eff-polarizedgowdy}}
    \label{table3}
    \begin{tabular}{c c c}
Method & $b_t$ & $m_t$ \\
\hline
deloused leapfrog & 4.8 & -2.1 \\
Crank-Nicholson & 9.6 & -2.8 
    \end{tabular}
    \end{table}

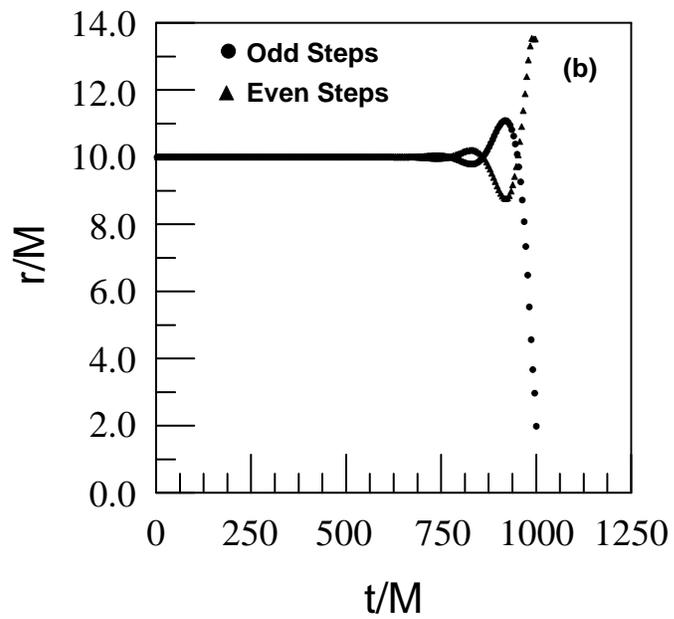
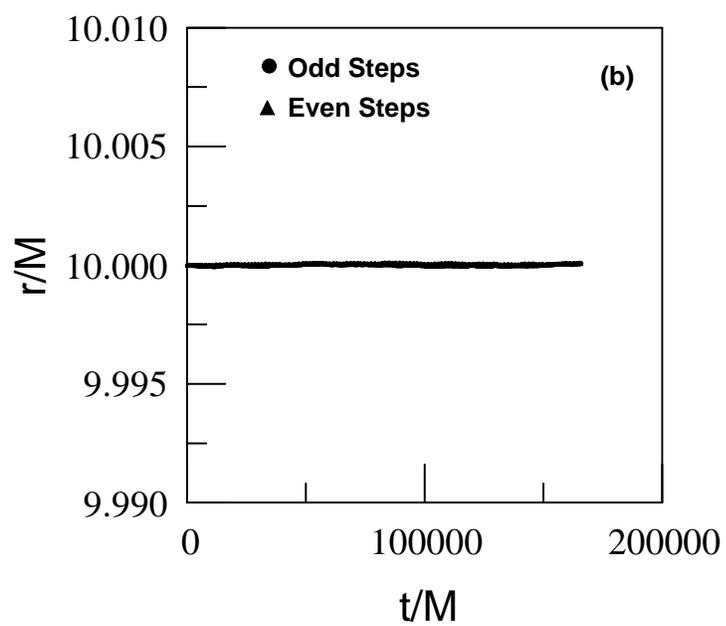
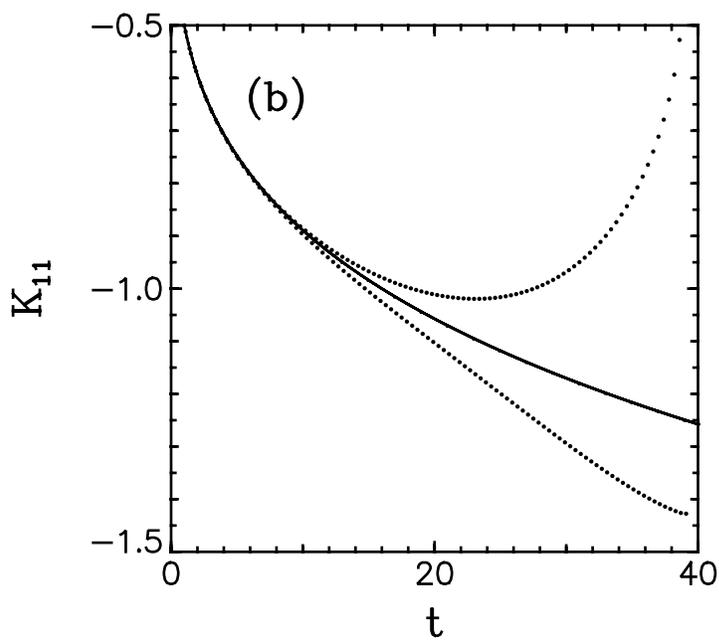
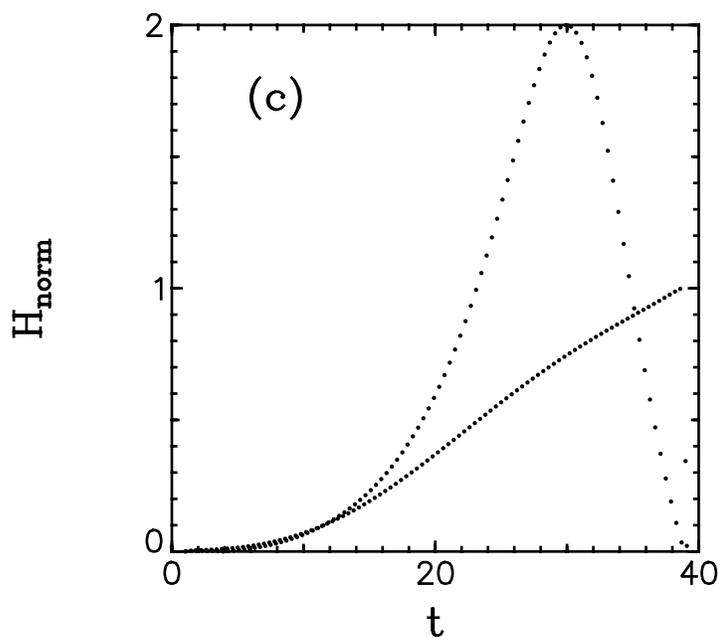
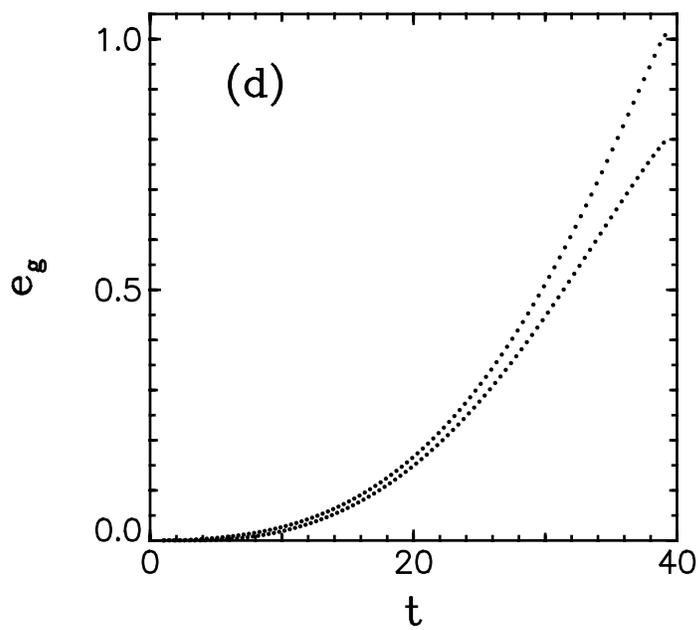
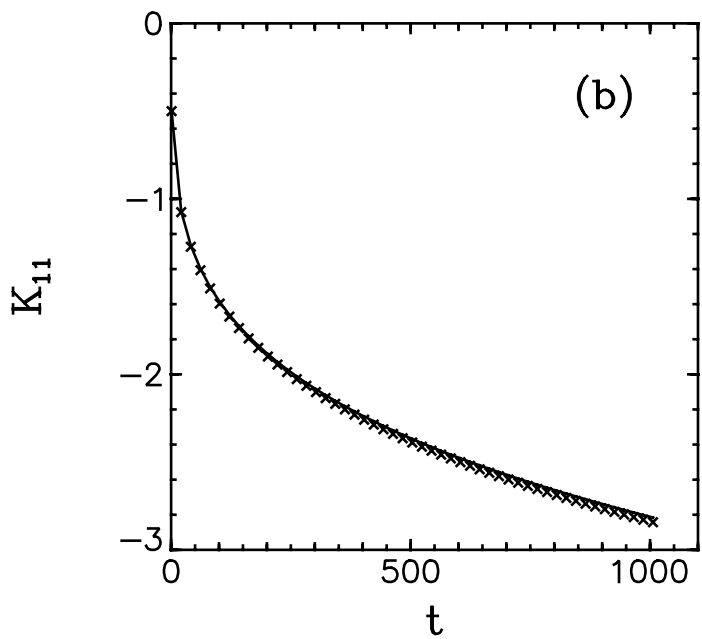
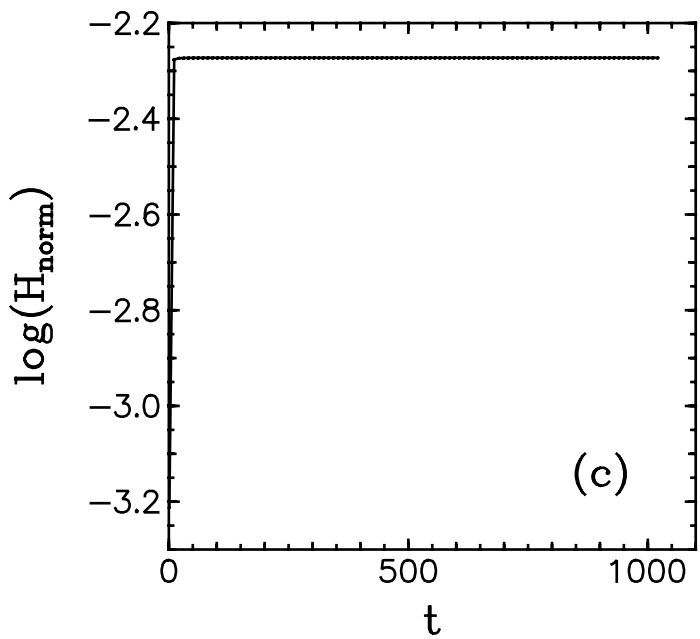
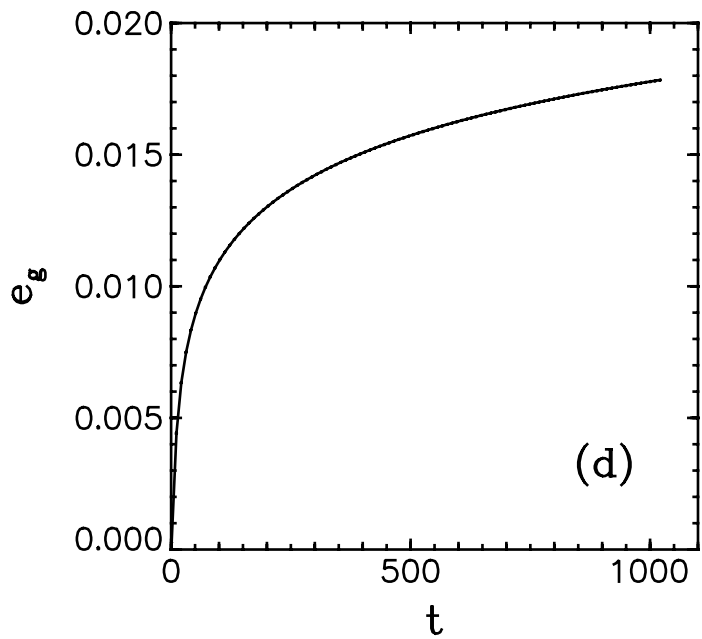
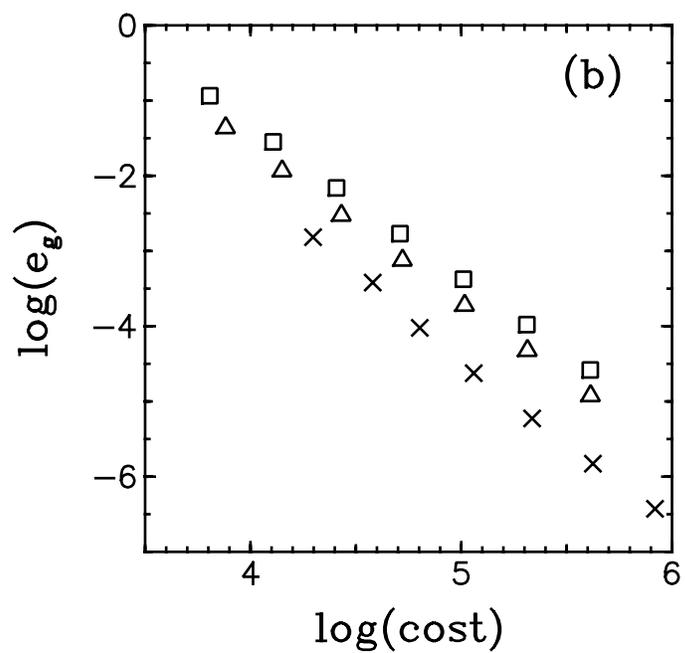
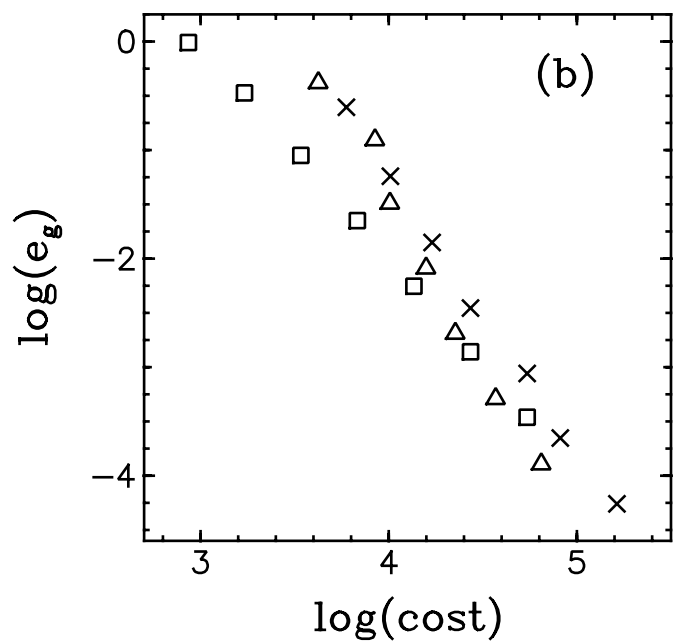
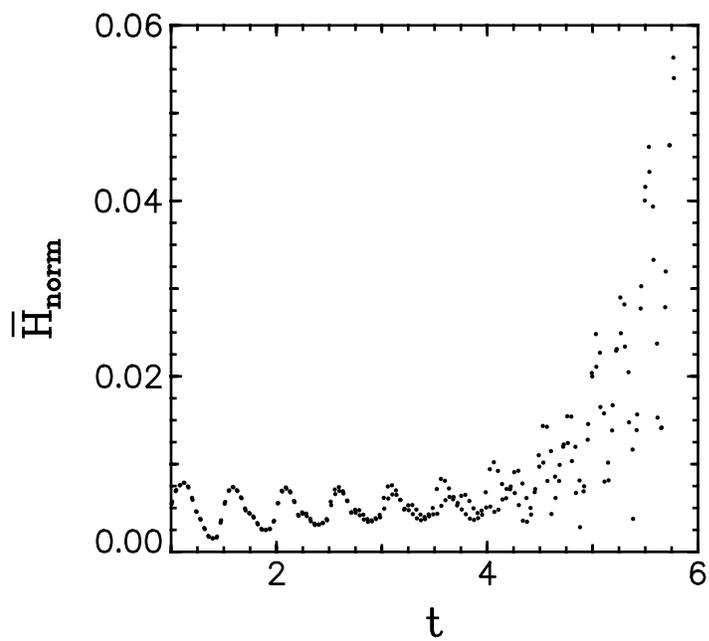
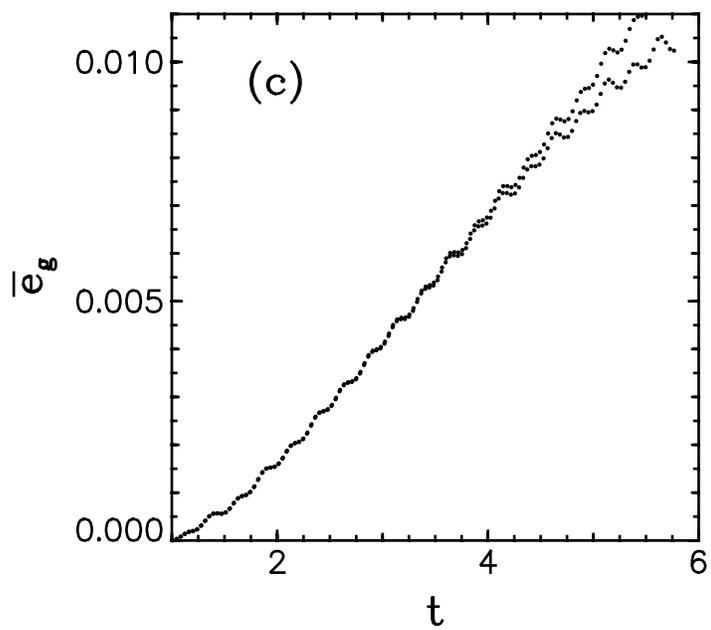
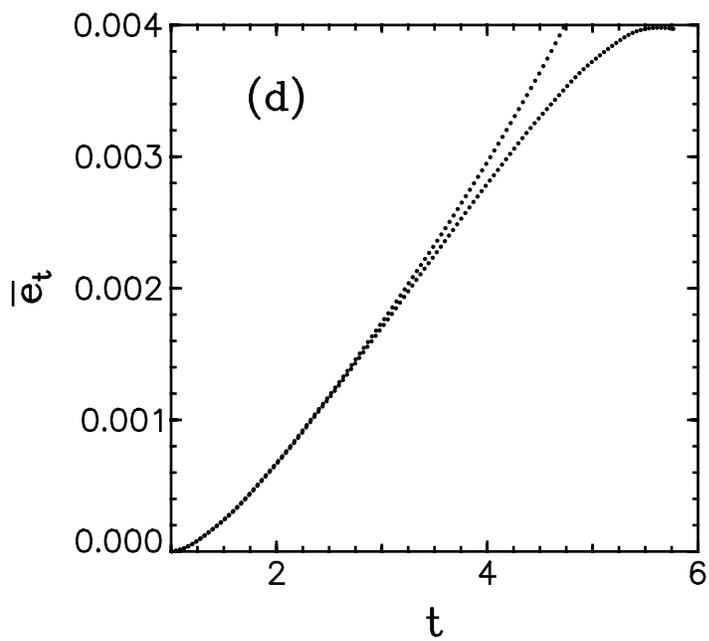
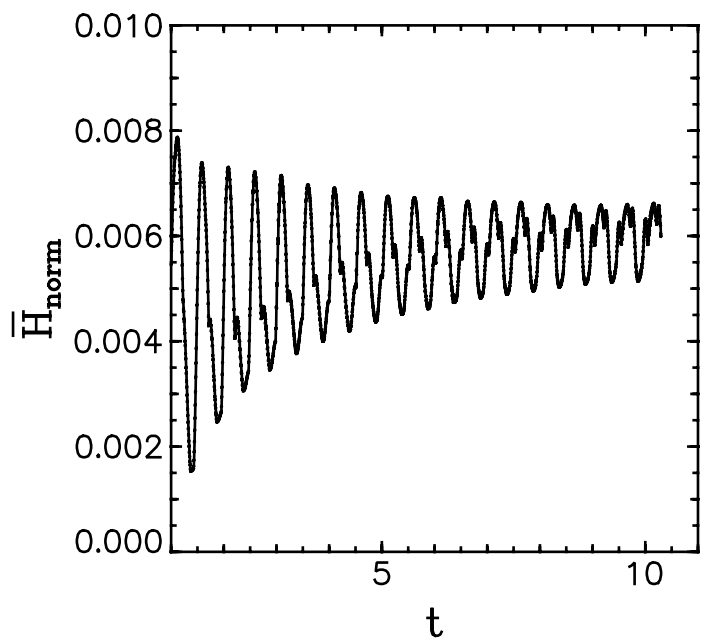
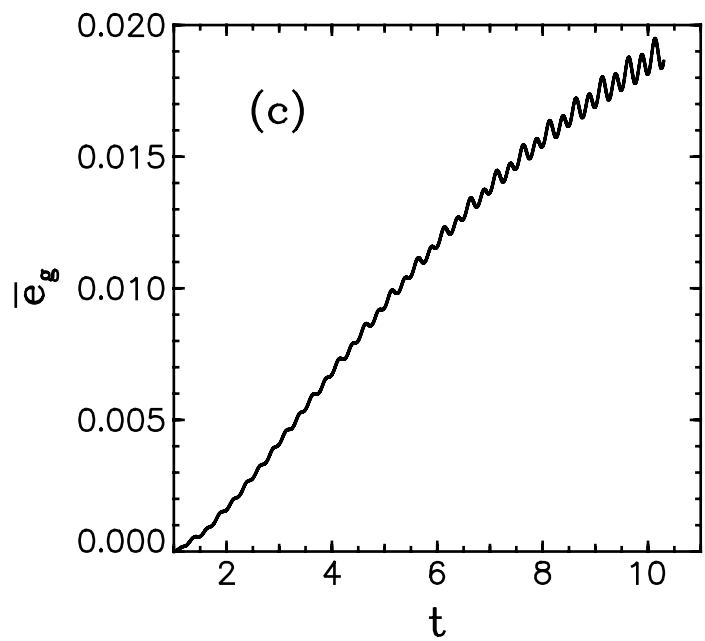
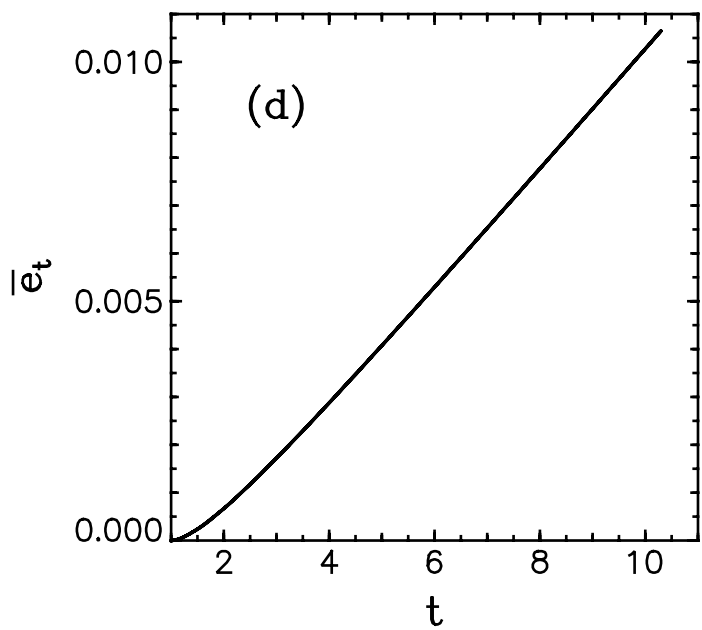
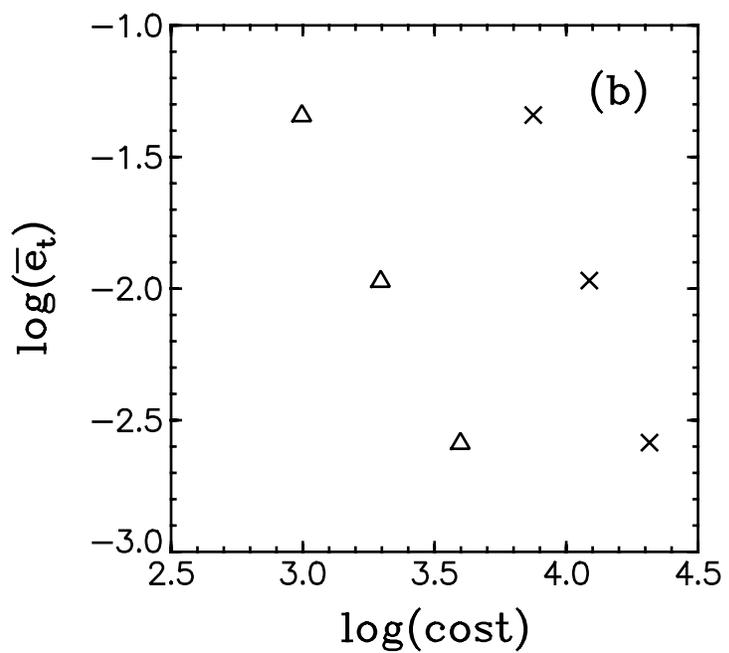
\end{document}